\documentclass[iop]{emulateapj}
  
\usepackage{apjfonts}
\usepackage{amsmath}
\usepackage{epsf}
\usepackage{color}
\usepackage{comment}
\usepackage{amsmath} 
\usepackage{multirow}

\shortauthors{Chiang, Overzier \& Gebhardt}

\newcommand{\lya}{Ly$\alpha$}

\begin{document}
\title{Ancient Light From Young Cosmic Cities: Physical and Observational Signatures of Galaxy Proto-clusters}

\author{Yi-Kuan Chiang\altaffilmark{1}, Roderik Overzier\altaffilmark{2, 1}, Karl Gebhardt\altaffilmark{1}}

\altaffiltext{1}{Department of Astronomy, University of Texas at Austin, 1 University Station C1400, Austin, TX 78712, USA}
\altaffiltext{2}{Observat\'orio Nacional, Rua Jos\'e Cristino, 77. CEP 20921-400, S\~ao Crist\'ov\~ao, Rio de Janeiro-RJ, Brazil}

\begin{abstract}
A growing number of galaxy clusters at $z=$1--2 is being discovered as part of deep optical, IR, X-ray, and Sunyaev--Zel'dovich effect surveys. For a complete picture of cluster formation, however, it is important that we also start probing the much earlier epoch, between redshifts of about 2 and 7, during which these clusters and their galaxies first began to form. Because the study of these so-called ``proto-clusters'' is currently quite limited by small number statistics, widely varying selection techniques, and many assumptions, we have performed a large systematic study of cluster formation utilizing cosmological simulations. We use the Millennium Simulations to track the evolution of dark matter and galaxies in about 3000 clusters from the earliest times to $z=0$. We define an effective radius $R_e$ for proto-clusters and characterize their growth in size and mass with cosmic time. We show that the progenitor regions of galaxy clusters (ranging in mass from $\sim10^{14}$ to a few times $10^{15}$ $M_\odot$) can already be identified in galaxy surveys at very early times (at least up to $z\sim5$), provided that the galaxy overdensities are measured on a sufficiently large-scale ($R_e\sim$5--10 Mpc comoving) and with sufficient statistics. We present the overdensities in matter, dark matter halos, and galaxies as functions of present-day cluster mass, redshift, bias, and window size that can be used to interpret the wide range of structures found in real surveys. We also derive the probability that a structure having a galaxy overdensity $\delta_{\textrm{gal}}$, defined by a set of observational selection criteria, is indeed a proto-cluster, and show how their $z=0$ masses can already be estimated long before virialization. Galaxy overdensity profiles as a function of radius are presented. We further show how the projected surface overdensities of proto-clusters decrease as the uncertainties in redshift measurements increase. We provide a table of proto-cluster candidates selected from the literature, and discuss their properties in the light of our simulations predictions. This paper provides the general framework that will allow us to extend the study of cluster formation out to much higher redshifts using the large number of proto-clusters that are expected to be discovered in, e.g., the upcoming HETDEX and Hyper Suprime-Cam surveys.

\end{abstract}

\section{Introduction}

Galaxy clusters are unique laboratories for both cosmology and galaxy formation, providing great leverage to test models for each. In the local universe, galaxy clusters are known to host well-established components such as the dominant red sequence galaxies, diffuse star light, and a hot X-ray emitting intracluster medium (ICM), which outweighs the total stellar mass. All of these are embedded in a virialized dark matter halo with a mass $\gtrsim$ 10$^{14}$ $M_{\odot}$. With diffuse X-ray, cluster red sequence, and Sunyaev--Zel'dovich effect searching techniques, massive clusters are found at $z < 1$ \citep[e.g.,][]{ebeling01, olsen07, foley11, menanteau12}, at $1 < z < 1.5$ \citep[e.g.,][]{gladders05, goto08, wilson09, fassbender11a} and $z\gtrsim1.5$ \citep[e.g.,][]{henry10, tanaka10, santos11}. However, these traditional techniques are all based on the presence of a prominent red sequence or ICM, and start to become biased toward the most virialized systems at $z\gtrsim1$. Furthermore, all these techniques reach their limit at $z\sim2$ due to the lack of mature cluster components.

It is important that we go to higher redshifts to probe the epoch of cluster formation. First of all, some of the most massive clusters at $z\gtrsim1$ appear remarkably mature, with a sufficiently deep potential well of dark matter halo, red sequence, and ICM in place, suggesting formation redshifts for most of their stellar contents much beyond 2 \citep{blakeslee03, mei06, andreon08, papovich10, rettura10, rettura11, fassbender11b}. Secondly, opposite to the morphology/star formation--density relation seen in the local universe \citep{dressler80, goto03}, galaxies in dense regions at high redshifts are found to experience enhanced star formation, interactions, and/or accelerated evolution, and active galactic nucleus (AGN) activity \citep{elbaz07, tran10, 2011MNRAS.418..938G, koyama13a, martini13} although a full consensus has not yet been reached. Thus, for a full census of cluster formation, it is important that we study them near the peak in the cosmic star formation and AGN activity at $z\gtrsim2$ \citep{hopkins06, fanidakis12}.

At $2\lesssim z \lesssim 7$, the so-called ``proto-clusters'' are predicted to have significant, large-scale overdensities of galaxies, allowing us to trace the evolution of clusters beyond the limit of traditional techniques. Some observational efforts have been made to search, identify, and characterize these proto-clusters with different techniques. Overdensities of narrow-band-selected emission line galaxies have been found around highly biased tracers like radio galaxies \citep{pentericci00, kurk00, kurk04a, venemans02, venemans04, venemans07, galametz10} and in ``random'' fields. Also, galaxy concentrations having a similarly narrow range in velocities have been discovered as part of spectroscopic follow-up of photo-$z$ or color-selected galaxies \citep{steidel98, steidel05, shimasaku03, shimasaku04, ouchi05, toshikawa12}. 

Although limited by small number statistics, these studies have revealed some intriguing properties of proto-clusters. For example, extended Ly$\alpha$ blobs are often found in proto-clusters \citep{prescott08, yang09, matsuda12}. Additionally, opposite to the strong suppression in star formation rate (SFR) seen in massive $z < 1.5$ clusters \citep{poggianti08, lidman08, patel09, bauer11}, proto-clusters show an excess of star-forming galaxies \citep[e.g.,][]{miley04, ouchi05, steidel05, overzier08, hayashi12}, extreme starbursts \citep{blain04, stevens03, capak11, ivison13}, and AGN activity \citep{pentericci02, croft05, lehmer09, digby-north10, martini13}. In some proto-clusters, the galaxies are on average older than field galaxies \citep{steidel05, hatch11b, koyama13a}, supporting the picture of accelerated galaxy formations in dense environments well before the clusters were fully formed. Recently, abnormal metallicities for proto-cluster galaxies have been reported \citep{kulas13}. Merging subclusters showing properties consistent with transitional stages between proto-clusters and clusters are also found \citep{gonzalez05, spitler12}.

A rich set of questions could be studied with proto-clusters. Are the abundance and growth of clusters consistent with the concordance cosmology and models of structure formation? How exactly do clusters assemble from the filamentary cosmic web, and how do they evolve from ``proto'' to mature? How do their galaxies become quenched in concordance with the red sequence and the morphology--density relation? However, it has been a challenge to answer such questions, mainly due to the observational and the intrinsic variety of the ``proto-cluster zoo'', combined with the fact that cluster progenitors, like clusters, are extremely rare.

One way in which this could be achieved is through a more thorough comparison between theoretical predictions and simulations. \citet[][]{steidel98} used analytical descriptions of structure formation theory to derive the total dark matter overdensity associated with one of the first proto-clusters discovered at $z\sim3$, allowing them to infer a total mass for the descendant cluster and its likely redshift of virialization (see also \citet{steidel05} for a similar case study at $z\sim2$).

In reality, the cluster formation process is much more complex as it depends on the hierarchical growth of dark matter and galaxies in three-dimensions on both large and small scales. A first step toward tracing the progenitor structures of galaxy clusters in $\Lambda$CDM was performed by \cite{suwa06}. They used cosmological ($N$-body) simulations to statistically quantify the overdensities in dark matter (halos) associated with clusters. They showed that some of the structures observed at high redshifts indeed have properties expected of proto-clusters, and derived probabilities for a given overdensity in dark matter to evolve into a cluster by $z=0$. \cite{angulo12} used an extremely large-volume simulation to study the evolution of the most massive halos at $z\sim6$, pointing out that the most massive halos do not always become the most massive clusters at $z=0$. The key factor that determines the final fate of high redshift halos is the surrounding matter overdensity on very large-scales, which is a much better indicator of its $z=0$ mass, $M_{z=0}$, than the halo mass by itself.

One complication of these kinds of theory--observation comparisons is the connection between the dark matter and the galaxies. \citet{saro09} ran hydro-dynamical simulations of cluster formation to compare the properties of galaxies and the ICM with those in a well-studied proto-cluster at $z=2.2$. \citet{DLB07} used the Millennium Run (MR) simulations and semi-analytic models of galaxy formation to predict the physical and observational properties of brightest cluster galaxies. An even closer match between simulations and observations can be achieved by constructing mock redshift surveys, and mimicking the various observational selection effects. For example, using this technique \cite{overzier09} was able to compare observations and simulations of the environments of quasi-stellar objects at $z\sim6$ as possible progenitor regions of galaxy clusters.

It has become clear that if we want to directly target the epoch of cluster formation, we need to develop reliable tools that can relate the main observables of proto-clusters to their main physical characteristics. Analogous to studies of galaxy formation, these tools need to be able to distinguish between structures of different masses, ages, and formation histories. In this paper, the first of a series of papers on the early formation history of galaxy clusters, we will present the characteristic properties for a sample of  $\sim3000$ galaxy proto-clusters in the MR. We study the statistical properties of overdensities in the distribution of dark matter, dark matter halos, and galaxies as a function of redshift, observational window size, and various halo and galaxy tracers.  By comparing with random regions, we derive the conditional probability that a structure with a given large-scale mass distribution is indeed a proto-cluster. We also show how the $z=0$ cluster mass can be estimated from the overdensity of galaxies, and how we could distinguish between progenitors of ``Fornax'', ``Virgo'', and ``Coma'' type clusters at redshifts as high as $\simeq2-5$. 

The structure of this paper is as follows. In Section 2, we describe our simulations-assisted approach and give our main definitions related to proto-clusters. In Section 3, we present the $\Lambda$CDM predictions for cluster assembly, size growth, overdensity evolution for fields and proto-clusters, as well as the identification and mass estimation for proto-clusters. In Section 4, we discuss our results and make a preliminary comparison with recent observations. We summarize our work in Section 5. If unspecified, the cosmological parameters used are based on $WMAP$1: $\Omega_m=0.25$, $\Omega_b=0.045$, $\Omega_{\Lambda}=0.75$, $h=0.73$, $n_s=1$, $\sigma_8=0.9$. $WMAP$7 cosmology is used when we present a comparison of the results for different cosmologies.

\section{Simulations and methods}

Although cluster formation is a continuing process throughout the history of the universe \citep[e.g.,][]{gonzalez05}, in this paper we will focus on $z>2$ as it marks the boundary between the epoch at $z<2$, in which we find the first observational evidence of large virialized clusters, and the epoch in which those first galaxy clusters are presumed to have been forming.

The need for a simulations-assisted approach is obvious. First, limited by the small number of observed proto-clusters and their varying selection techniques, it is premature to constrain models using these results. However, based on observations of the local universe and the initial conditions imprinted in the cosmic microwave background, sophisticated cosmological simulations have been able to make highly physically motivated predictions for the high redshift universe. Second, with the coming of future redshift surveys which will provide detailed information with large statistics, it becomes more and more important to extract equally detailed predictions from simulations in order to understand the non-trivial relations between model parameters and observables. Finally, predictions from simulations are crucial to help design observations and optimize analysis techniques. Here we describe the simulations and our methodology in this work.

\subsection{Cosmological $N$-body Simulations and Semi-analytic Galaxy Catalogs}
To study the high redshift progenitors of the most extreme present-day structures and their galaxy contents, we require simulations that span an enormous range in physical length scales, masses, and redshifts, and also have a fine treatment of the most important baryonic processes. Therefore, in this work we use the MR dark matter $N$-body simulation \citep{springel05} and a recent semi-analytic galaxy formation model \citep{guo11}. For detailed descriptions of the methods and implementation, we refer the reader to previous works by the MR group \citep[e.g.,][]{springel05, croton06, lemson06, DLB07, guo11}. 

The MR simulation gravitationally evolved $2160^3$ dark matter particles with mass $8.6\times 10^8$ $M_{\odot}$ $h^{-1}$ in a comoving box of 500 Mpc $h^{-1}$ on a side, from $z=127$ to $z=0$. The original (2005) run used a $\Lambda$CDM universe with $\Omega_m=0.25$, $\Omega_b=0.045$, $\Omega_{\Lambda}=0.75$, $h=0.73$, $n_s=1$, $\sigma_8=0.9$, based on the combined analysis of the 2dF Galaxy Redshift Survey \citep{colless01} and the first year $Wilkinson\ Microwave\ Anisotropy\ Probe$ \citep[$WMAP$;][]{spergel03}. The dark matter distribution was processed by the standard friends-of-friends (FOF) group finder and the subhalo finder, SUBFIND \citep{springel01}, at 64 discrete epochs. The field dark matter density and the subhalo catalog were stored for these 64 "snapshots". The merger trees were then constructed by identifying and linking the progenitors and descendants. Recently, a new run based on the $WMAP$7 cosmology \citep{komatsu11} was released \citep{guo13}. In this paper, we will primarily use the $WMAP$1 run due to the availability of low level data products such as the original particle density field. We also present the $WMAP$7 results as comparison, showing that the results using these two cosmologies are quantitatively similar.

The semi-analytic galaxy model (SAM) simulates galaxy formation based on subhalo merger trees. Galaxies are formed in the subhalos and interact hierarchically. They gain stars through local star formation within an assumed interstellar medium and through merger/accretion events. The basic recipes include gas cooling and infall, reionization heating, black hole growth, AGN and supernova feedback, and a realistic gradual gas stripping process that operates when galaxies become satellites. The free parameters in the model are then determined by matching with the observed galaxy abundance as a function of galaxy properties in the local universe. Recently, a $WMAP$7 version was released \citep{guo13}.

The results of these simulations have been widely used and compared with observations of various aspects of the galaxy population \citep[e.g.,][]{croton06, cohn07, genel08, genel09, bertone09, overzier09, guo09, guo11, horesh11, bahe12, henriques12, quilis12, merson13}, finding that galaxy properties and the large-scale clustering are reasonably well-reproduced from low to high redshifts. Specifically, \cite{guo11} have shown that for galaxy clusters in the local universe, the cluster abundance, cluster galaxy luminosity function, and galaxy number density profiles match very well with those found in large surveys such as the Sloan Digital Sky Survey. Interestingly, they also predict the existence of the so called ``orphan galaxies'' in clusters with their dark matter subhalos being stripped below the mass resolution of the MR. At $1.5 \lesssim z \lesssim 3$, \cite{guo09} have shown that the abundances, redshift distributions, clustering, and SFRs of Lyman break galaxies (LBGs), star-forming galaxies (BXs), and distant red galaxies are basically reproduced \citep[see also][for a comparison of the Durham galaxy model with observations for BzK-selected galaxies]{merson13}. \cite{henriques12} and \cite{overzier09} further push the comparison out to redshifts 4 and 6, respectively. In our work, we use the models to link galaxy to halo and eventually mass overdensity, thus only making the assumption that model galaxies form in the right halos, and that the halos have the correct clustering. In this paper, we do not use any photometric predictions of galaxies, as this will add significant additional uncertainties and model assumptions. In future work, however, we will use the Millennium Run Observatory \citep{overzier13}, a virtual observatory framework built on top of the Millennium SAM galaxy catalog, in order to more realistically compare simulations with observations.


\begin{figure}[]
\epsscale{1.1}
\plotone{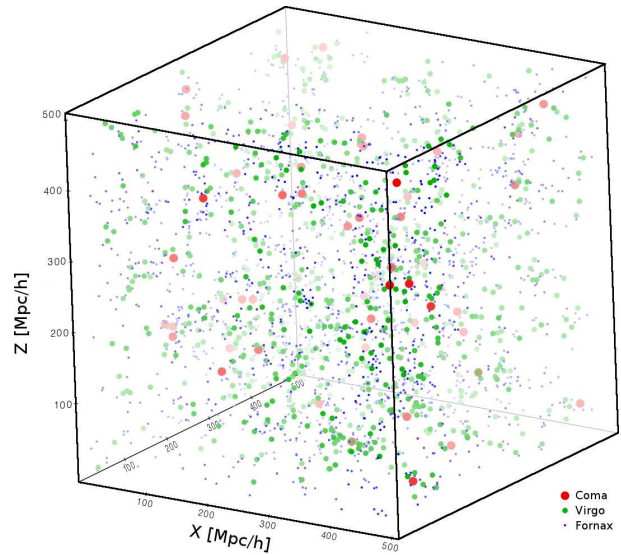}
\vspace{2mm}
\caption{Spatial distribution of the 2832 galaxy clusters at $z=0$ in the MR Simulation, with masses of 1.37--$3 \times$ (blue), 3--$10 \times$ (green) and $>10 \times$ (red) $10^{14}$ $M_{\odot}$. The analysis performed in this paper is based on this cluster sample.}
\end{figure}

\begin{figure*}[!ht]
\epsscale{1.17}
\plottwo{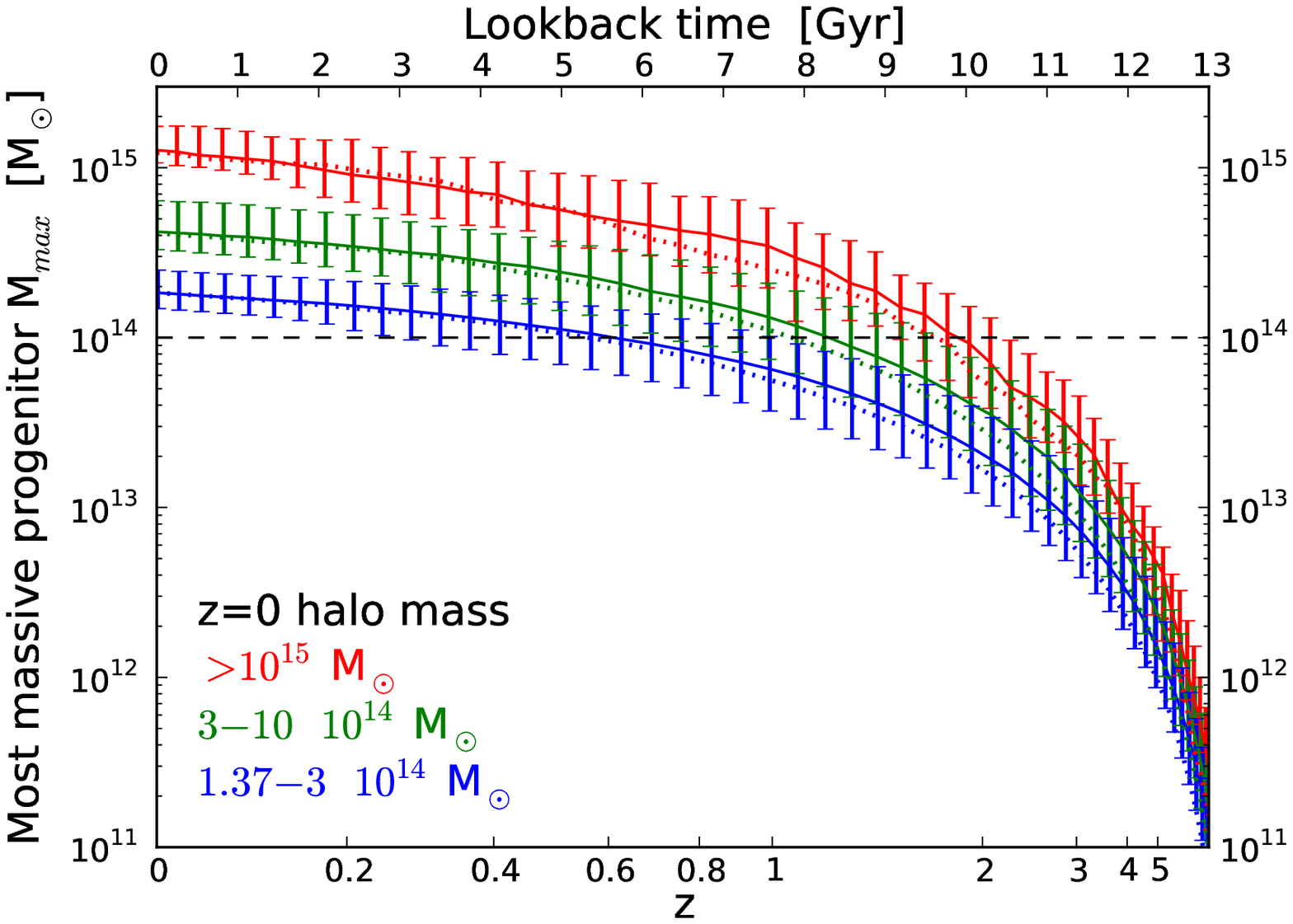}{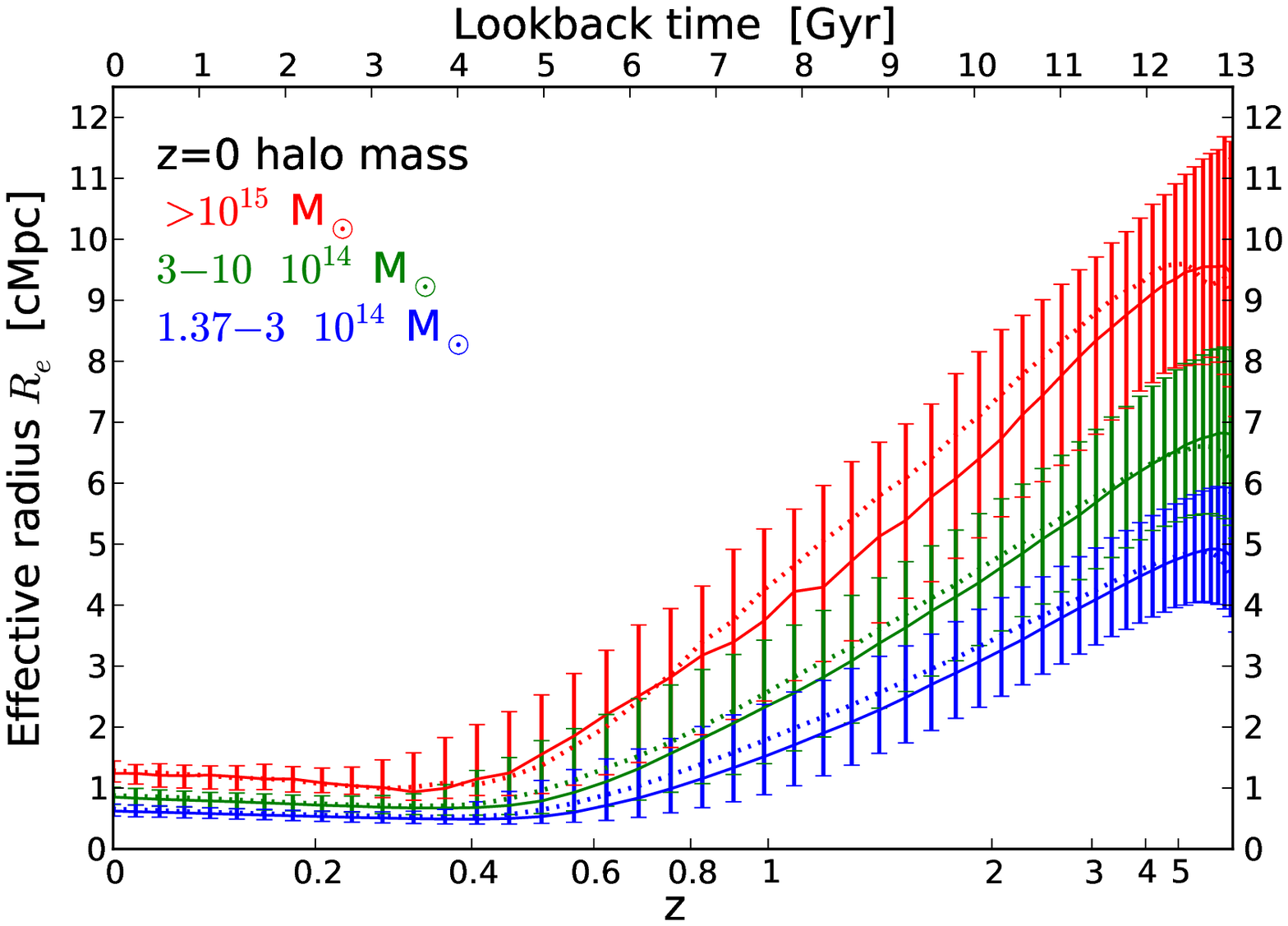}
\caption{Evolution of the mass of the most massive progenitor halo $M_{\textrm{max}}$ (left) and the effective radius $R_e$ (right) of (proto-)clusters binned by $M_{z=0}$. Results based on $WMAP$1 (solid) and $WMAP$7 (dotted) cosmologies are presented and matched by redshift. The lines and error bars indicate the medians and $1-\sigma$ scatter (15.865th and 84.135th percentile). We note that the sizes of proto-clusters evolved largely before $z \sim 2$, and more massive proto-clusters occupied larger comoving volumes and formed cluster-size halos earlier. }
\end{figure*}

\subsection{Definitions of Clusters and Proto-clusters}
Observationally, galaxy clusters are identified by overdensities of galaxies, dark matter, and hot ICM on scales of about 1 Mpc. Galaxy kinematics and gravitational lensing studies show that cluster galaxies are embedded in a massive halo of dark matter with mass of $\gtrsim 10^{14}$ $M_{\odot}$, which weighs about five times more than its baryonic contents \citep{gonzalez07, giodini09, andreon10}. It is a common belief that the dark matter halo mass is the most important underlying property when distinguishing between clusters, groups, and field galaxies. All other features of galaxy clusters can, to the first order, be treated as manifestations of halo mass and assembly history. 

For these reasons, in the simulations we will define a galaxy cluster as a gravitationally bound and virialized dark matter halo (and its associated galaxies) with halo mass $> 10^{14}$ $M_{\odot}$ $h^{-1}$.\footnote{We use the ``$m_{\textrm{tophat}}$'' defined by the mass within the radius where the halo has an overdensity equal to the spherical collapse model threshold in our cosmology. The ``$m_{\textrm{tophat}}$'' for a given cluster is, on average, higher than the ``$m_{\textrm{crit200}}$'' by $\sim 25\%$.} Then, a proto-cluster is defined simply as the high redshift progenitor of such a galaxy cluster at $z=0$.\footnote{We walk the merger trees which trace progenitors of all the identified and disrupted subhalos within the FOF group of a $z=0$ cluster.} It is important to note that for most practical purposes, the term proto-cluster does not refer to a single virialized object, but rather to a large region in space containing all the dark matter and baryons that merge into one massive bound virialized structure by $z=0$. By tracing the halo merger trees of all $z=0$ clusters in the MR simulation, we are able to identify proto-clusters at high redshifts and quantify their properties predicted in a $\Lambda$CDM universe. Based on these definitions, we compile a sample of 2832 galaxy clusters with a halo mass of $> 10^{14}$ $M_{\odot}$ $h^{-1}$ at $z=0$. This sample consists of 1976 low-mass ``Fornax-type'' clusters of 1.37--$3\times 10^{14}$ $M_{\odot}$, 797 intermediate mass ``Virgo-type'' clusters of 3--$10 \times 10^{14}$ $M_{\odot}$, and 59 high-mass ``Coma-type'' clusters of $> 10^{15}$ $M_{\odot}$. The three-dimensional locations of the clusters in the MR are indicated in Figure. 1.

\subsection{Overdensity}
Because of the hierarchical nature of the structure formation, galaxy clusters formed in regions with the largest initial overdensities all the way from small to large-scales. Therefore, the progenitors of galaxy clusters have manifested themselves since the earliest time by overdensities of dark matter, halo number, and galaxy number within a certain volume. We use the common definitions of matter, halo, and galaxy overdensities as follows:
\begin{equation}
\delta_m(\textit{\textbf{x}})\equiv\frac{\rho(\textit{\textbf{x}})-\langle\rho\rangle}{\langle\rho\rangle},
\end{equation}
\begin{equation}
\delta_h(\textit{\textbf{x}})\equiv\frac{n_h(\textit{\textbf{x}})-\langle n_h\rangle}{\langle n_h\rangle},
\end{equation}
\begin{equation}
\delta_{\textrm{gal}}(\textit{\textbf{x}})\equiv\frac{n_{\textrm{gal}}(\textit{\textbf{x}})-\langle n_{\textrm{gal}}\rangle}{\langle n_{\textrm{gal}}\rangle},
\end{equation}
where $\delta_m(\textit{\textbf{x}})$, $\delta_h(\textit{\textbf{x}})$ and $\delta_{\textrm{gal}}(\textit{\textbf{x}})$ are overdensities of dark matter, halo, and galaxy number, respectively. $\langle\rho\rangle$, $\langle n_h\rangle$, and $\langle n_{\textrm{gal}}\rangle$ are the ensemble averages of the density of dark matter, halo, and galaxy number, respectively. The Ergodic principle is applied to calculate these values. When calculating $\delta_h$ and $\delta_{\textrm{gal}}$, properties of halos (usually mass) and galaxies (e.g., stellar mass and SFR) need to be specified. In addition, these overdensities are typically calculated in windows with different sizes, shapes, and weighting profiles. In this work, we use tophat weighted cubic windows of various sizes in comoving coordinates. These windows should be large, ranging from a few to several tens of Mpc, in appreciation of the fact that galaxy proto-clusters are generally very large.

\section{Results}
In this section we present $\Lambda$CDM predicted properties of proto-clusters extracted from the MR simulation and SAM galaxy catalog. In the previous section, we identified 2832 galaxy clusters with halo mass $>10^{14}$ $M_{\odot}$ $h^{-1}$ at $z=0$ (see Figure 1). Here we trace their progenitors to high redshifts. The center of a proto-cluster is defined as the center of mass of its member halos. To relate the structures to observables, we quantify the key features such as mass, size, assembly history, overdensities of dark matter, halo, and galaxies. By comparing proto-cluster regions with random fields, we are able to statistically classify structures given a set of observed features. The length scales presented here are in comoving units.

\subsection{Assembly and Virialization Redshifts}
Under our definition, the birth of a galaxy cluster can be dated to the redshift at which the mass of the main halo first exceeded $\sim 10^{14}$ $M_{\odot}$. Further relaxation and virialization may be achieved after about one dynamical time, which is about $10^9$ yr, an order of magnitude shorter than the Hubble time. In Figure 2 (left), we show the most massive progenitor halo mass, $M_{\textrm{max}}$, of our MR cluster sample binned by $M_{z=0}$ as a function of redshift. Results based on $WMAP$1 (solid) and $WMAP$7 (dotted) cosmologies are presented. Medians and $1-\sigma$ scatter (15.865th and 84.135th percentile) of each bin are plotted. In general, massive clusters form earlier. The assembly redshifts for ``Coma'' ($>10^{15}$ $M_{\odot}$), ``Virgo'' (3--$10\times 10^{14}$ $M_{\odot}$), and ``Fornax'' (1.37--$3\times 10^{14}$ $M_{\odot}$) type clusters are about 1.5--2.3, 0.7--1.6 and 0.2--1, respectively. That is to say, the first objects that reached the ``threshold'' cluster mass of $10^{14}$ $M_{\odot}$ were likely the progenitors of ``Coma-type'' and more massive clusters at around $z=2.3$, while low-mass clusters made the transition from proto-cluster to cluster much more recently. The high value of $\sigma_8$ and low value of $\Omega_m$ in the $WMAP$1 run partially compensate each other \citep[demonstrated in][]{guo13}, such that clusters form at only slightly higher redshifts ($< 10 \%$) compared to the $WMAP$7 run. Our results of cluster halo mass growth are consistent with those found by the RHAPSODY cluster simulations \citep{wu13}.

We note that cluster formation and halo assembly are ongoing processes. According to Figure 2 (left), to keep increasing the mass, a high redshift proto-cluster must be surrounded by many smaller halos waiting to be assembled onto the main halos. The sizes of these extended structures are quantified in the right panel of Figure 2 (see the next subsection for details). 
\subsection{Spatial Evolution of Proto-clusters}
At the redshifts prior to that at which a cluster assembles most of its mass into one single structure, a proto-cluster consists of many halos separated within a much larger volume. To quantify the spatial distribution and the size of the whole structure, we introduce, for the first time, an effective radius $R_e$ of proto-clusters. We define $R_e$ by the second moment of the member halo positions weighted by halo mass,
\begin{equation}
R_e \equiv \sqrt{\frac{1}{M}\Sigma_i m_i (\textbf{\textit{x}}_{\textbf{\textit{i}}}-\textbf{\textit{x}}_{\textbf{\textit{c}}})^2},
\end{equation}
where $M$ is the total mass of the proto-cluster in bound halos at the redshift of interest, $m_i$ is the mass of each halo and $\textbf{\textit{x}}_{\textbf{\textit{i}}}$ and $\textbf{\textit{x}}_{\textbf{\textit{c}}}$ are the position of each halo and the center of mass for the whole proto-cluster, respectively. We set a lower limit of $R_e$ to the half mass radius of the main cluster halo since once a cluster finished its assembly into a single halo, Equation (4) gives a spatial dispersion of zero. In this case, the size of the main halo properly indicates the scale at which the mass is distributed.

We note that in our definition $R_e$ is fundamental and SAM independent. $R_e$ is representative of the proto-cluster size, in a sense that a significant fraction of mass is within this radius. At $2 \lesssim z \lesssim 5$, about $65 \%$ of the mass in bound halos and $40 \%$ of the total mass of the proto-cluster is distributed inside $R_e$, and these fractions are independent of cluster mass. Later in the paper, we will use these fractions to construct a way to estimate cluster mass observationally using overdensity and effective volume. The defined $R_e$ is not sensitive to outliers or the likely departure from spherical symmetry. Figure 2 (right) shows $R_e$ for (proto-)clusters as a function of redshift binned by $M_{z=0}$. Results based on $WMAP$1 (solid) and $WMAP$7 (dotted) cosmologies are presented. The lines and error bars indicate the medians and $1-\sigma$ scatter (15.865th and 84.135th percentile) for each bin. We note that the sizes of proto-clusters evolved largely in the past. As expected, more massive proto-clusters occupied larger comoving volumes. The effective diameter $2R_e$ at $z\sim 2$ for a ``Coma'', ``Virgo'', and ``Fornax'' type proto-cluster is expected to be $13.0^{+3.8}_{-2.6}$, $9.0^{+2.4}_{-2.2}$ and $6.4^{+1.8}_{-1.6}$ Mpc (comoving), respectively. At $z\sim 5$, these sizes increase to be about $18.8^{+3.2}_{-3.2}$, $13.2^{+2.8}_{-2.4}$ and $9.6^{+1.8}_{-1.6}$ Mpc (comoving). Another way to look at the effective radius $R_e$ is that $R_e$ is very close to the radius of the Lagrangian volume of a halo in the simple tophat spherical collapse model. For example, a typical ``Virgo-type'' cluster reaches $10^{14}$ $M_{\odot}$ at $z \sim 1$ (Figure 2, left). Assuming this halo is formed under the growth of a tophat density perturbation, the radius of this overdense region is $\sim 6$ cMpc at $z=3$, which agrees well with the $R_e$ shown in Figure 2 (right). The high $\sigma_8$ and low $\Omega_m$ in the $WMAP$1 run leads to slightly smaller proto-clusters compared to those in the $WMAP$7 run at a given redshift. This is a direct consequence of the slightly higher cluster formation redshifts in $WMAP$1. However, the differences between the two cosmologies are only at a level of few percent. Since this characteristic radius does not contain the entire proto-cluster mass, the overdensity associated with a proto-cluster often extends even farther (see next subsection).

\begin{figure}[]
\epsscale{1.17}
\plotone{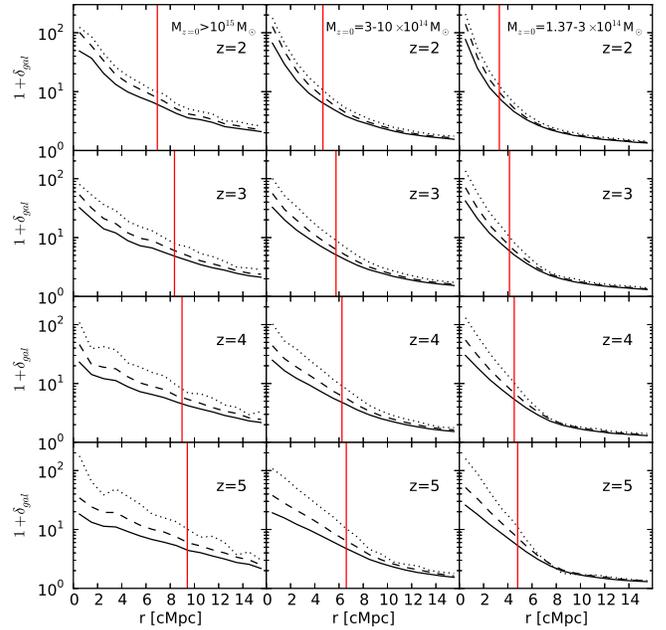}
\caption{Stacked differential overdensity profiles of proto-cluster galaxies in three present-day cluster mass bins (left to right) at redshifts 2, 3, 4, and 5 (top to bottom). Galaxies with star formation rate $>1$ $M_{\odot}$ yr$^{-1}$, stellar mass $>10^9$, and $>10^{10}$ $M_{\odot}$ are shown in solid, dashed, and dotted lines, respectively. Red lines indicate the effective radius $R_e$ defined in Section 3.2 for each subsample and redshift.}
\end{figure}

\subsection{ Overdensity Profile of Proto-cluster Galaxies}
To further demonstrate the large sizes of proto-clusters, as well as the detailed spatial distribution of proto-cluster galaxies as a function of cluster mass and redshift, we perform a stacking analysis of regions centered at proto-clusters. Figure 3 shows the stacked overdensity profiles of three populations of galaxies in three present-day mass bins (left to right) at redshifts 2, 3, 4, and 5 (top to bottom). The effective radius $R_e$ defined in Section 3.2 for each subsample and redshift is shown with red lines, showing how the observed density profiles are linked to the more fundamental and relatively model-independent $R_e$. In all cases, the density profile shows a steeper increase toward the center for more biased galaxy populations, exactly as expected. More massive galaxies (dotted lines) result in larger overdensities compared to less massive galaxies (dashed lines). Galaxies selected on the basis of a SFR $>1$ $M_{\odot}$ yr$^{-1}$ result in the lowest overdensities. The implications of using these different tracer populations will be discussed below. The galaxy overdensities that will be presented throughout this paper correspond to the values measured when integrating these density profiles over some given volumes.

Although individual structures show certain degrees of non-spherical symmetry and complicated topology associated with cosmic filaments, these averaged profiles are illustrative of the overall large sizes of proto-clusters that can be compared with real observations. We also note that some galaxies at the outskirts, although associated with the proto-cluster overdensities, will not become cluster members by $z=0$, but in the future, in appreciation of the fact that cluster formation is an ongoing and inside-out hierarchical process.

\begin{figure}[]
\epsscale{1.2}
\plotone{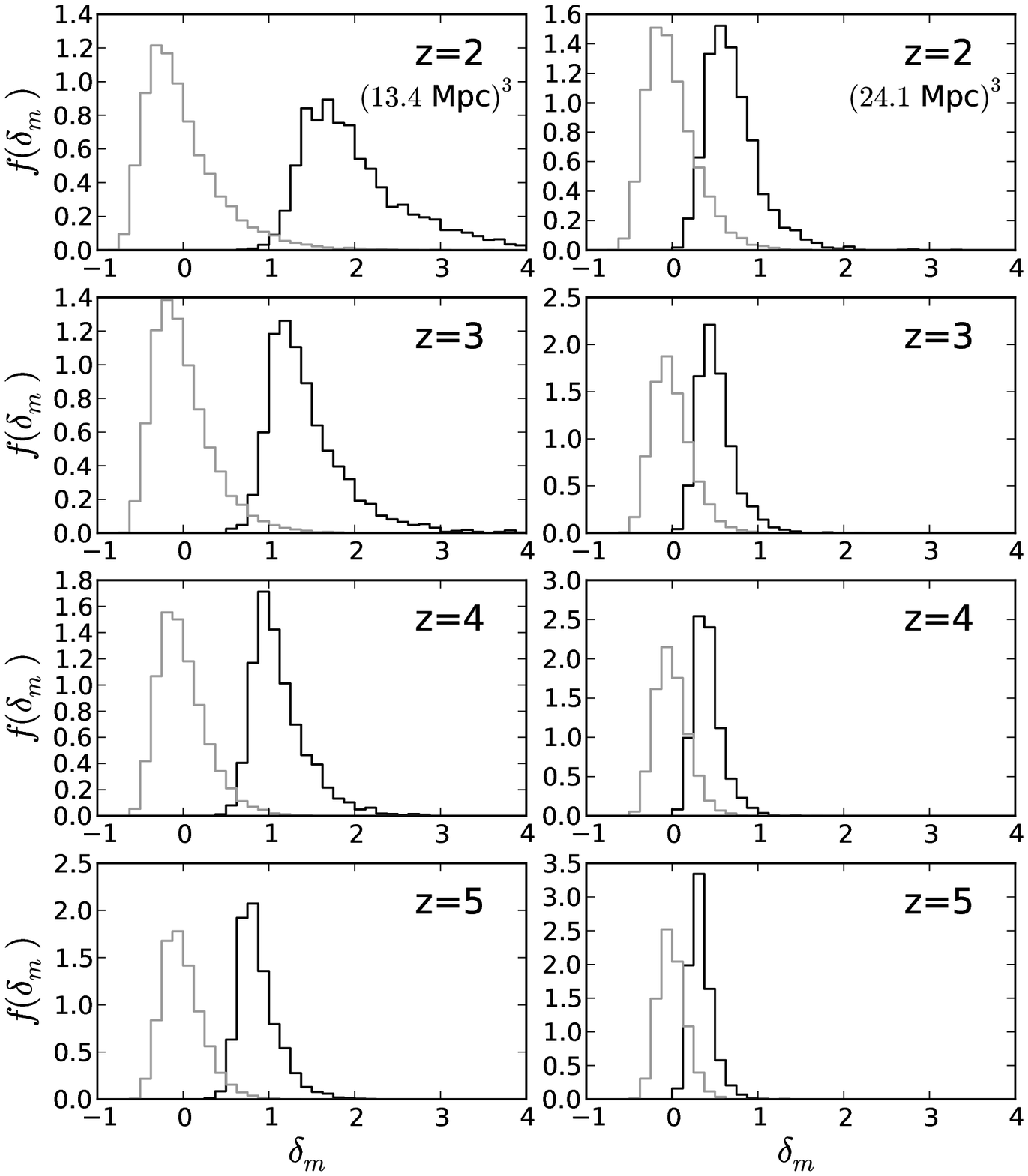}
\plotone{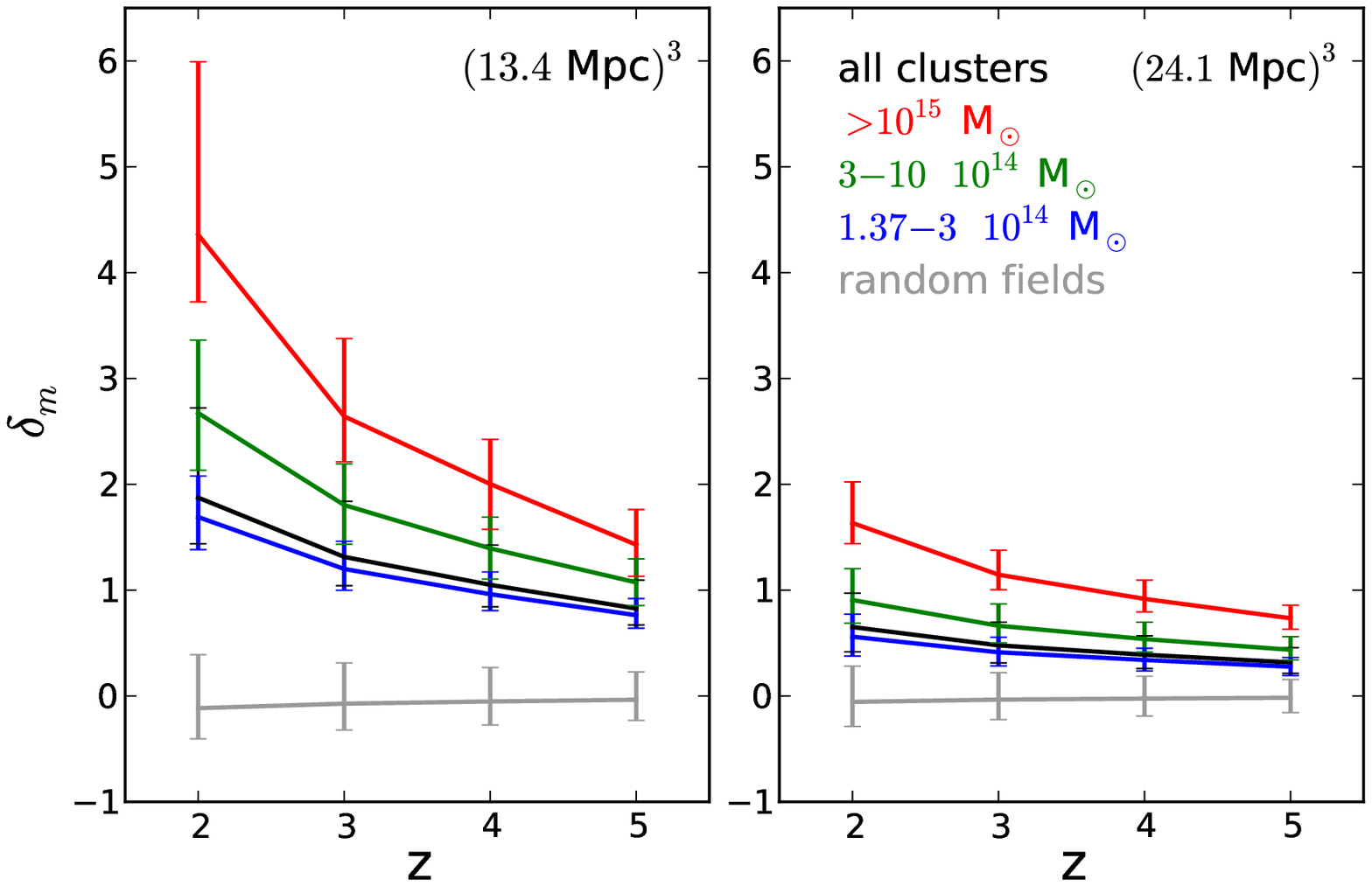}
\caption{Mass overdensities $\delta_m$ of proto-cluster regions (black) and 10,000 random regions (gray) at redshifts 2, 3, 4, and 5 calculated with (13.4 Mpc)$^3$ (left) and (24.1 Mpc)$^3$ (right) comoving tophat box windows. The top eight panels show the probability density distributions, $f(\delta_m)$ vs. $\delta_m$. The bottom two panels show the median $\delta_m$ and $1-\sigma$ scatter, with proto-cluster regions further binned by $M_{z=0}$. Proto-clusters can be recognized by high $\delta_m$. A larger window is better for separating massive proto-clusters from lower mass ones especially at high redshifts, in the sense that the scatter and overlap in $\delta_m$ are reduced. This is simply because of the sizes of proto-clusters shown in Figure 2.} 
\end{figure}

\subsection{Mass Overdensity}

The basic physical property determining the fate of structures is the initial density contrast of dark matter as a function of scale shortly after recombination. This density contrast then evolves under gravitational contraction. Therefore, we expect that the present-day mass of the structures is closely related to the overdensity of dark matter at high redshift averaged over appropriate volumes.

We test this scenario and quantify the correlation and scatter using our large cluster sample and 10,000 random regions drawn from the MR simulation. In the top eight panels of Figure 4, we plot the probability density distribution of dark matter overdensity, $\delta_m$ for (proto-)clusters (black) and random regions (gray) at $z=2,3,4,5$. The $\delta_m$ in left and right panels are calculated using tophat cubic windows with 13.4 Mpc and 24.1 Mpc comoving on a side, respectively. The probability density distributions here by definition are normalized to $1$ when integrating from $-1$ to $+\infty$. Similarly, for the bottom panels, we plot the medians and $1-\sigma$ scatter of these distributions for random regions (gray), all (proto-)clusters (black), and also (proto-)clusters binned by $M_{z=0}$ (red, green and blue) as functions of redshift. The high $\delta_m$ tails of the random regions may cover parts of nearby cluster or group regions. It is clear that proto-cluster regions have a higher $\delta_m$, and thus they stand out from random fields in overdensity space at all redshifts. Therefore, if we can infer $\delta_m$ from observables, we should be able to identify proto-clusters long before virialization, and even pin down their approximate $z=0$ masses. Also shown here is that $\delta_m$ increases with cosmic time for initially overdense regions. For the random regions, the medians slightly decrease with time and the scatter increase. This is consistent with the picture of structure formation that ``the rich get richer'' and the voids become emptier. Since in the nonlinear regime of structure formation, most of the volume in the universe is underdense, the median $\delta_m$ of random regions drops.

The relatively large windows we used are motivated by the sizes of proto-clusters presented previously. In smaller windows, $\delta_m$ are naturally higher since matter is clustered and the center of proto-clusters are close to the local maximum of $\delta_m$. However, what we really need is to choose an appropriate window to maximize the ability to distinguish the structures of interest. In general, larger windows are better for more massive clusters and at higher redshifts (recall Figure 2). In Figure 4, 13.4 Mpc windows do a better job in separating low-mass proto-clusters from the fields in the $\delta_m$ distribution at all redshifts. But the most massive proto-clusters stand out from the field and lower mass clusters by using the larger 24.1 Mpc window. This can be seen in the bottom right panel: the red line is completely separated from the green line without overlap of their scatter. Larger windows make the fractional-scatters of $\delta_m$ of massive proto-clusters smaller.

\begin{figure}[]
\epsscale{1.2}
\plotone{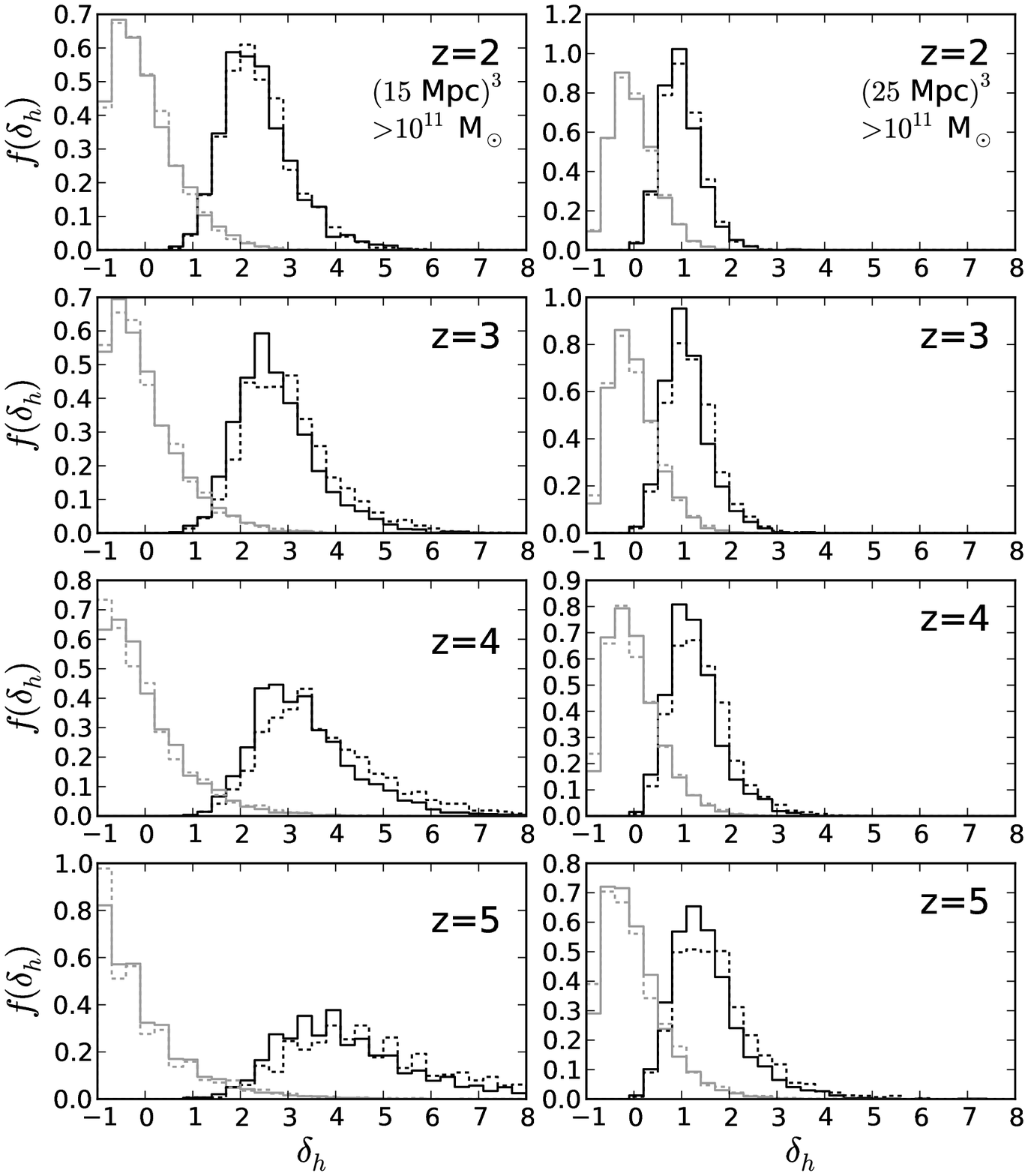}
\plotone{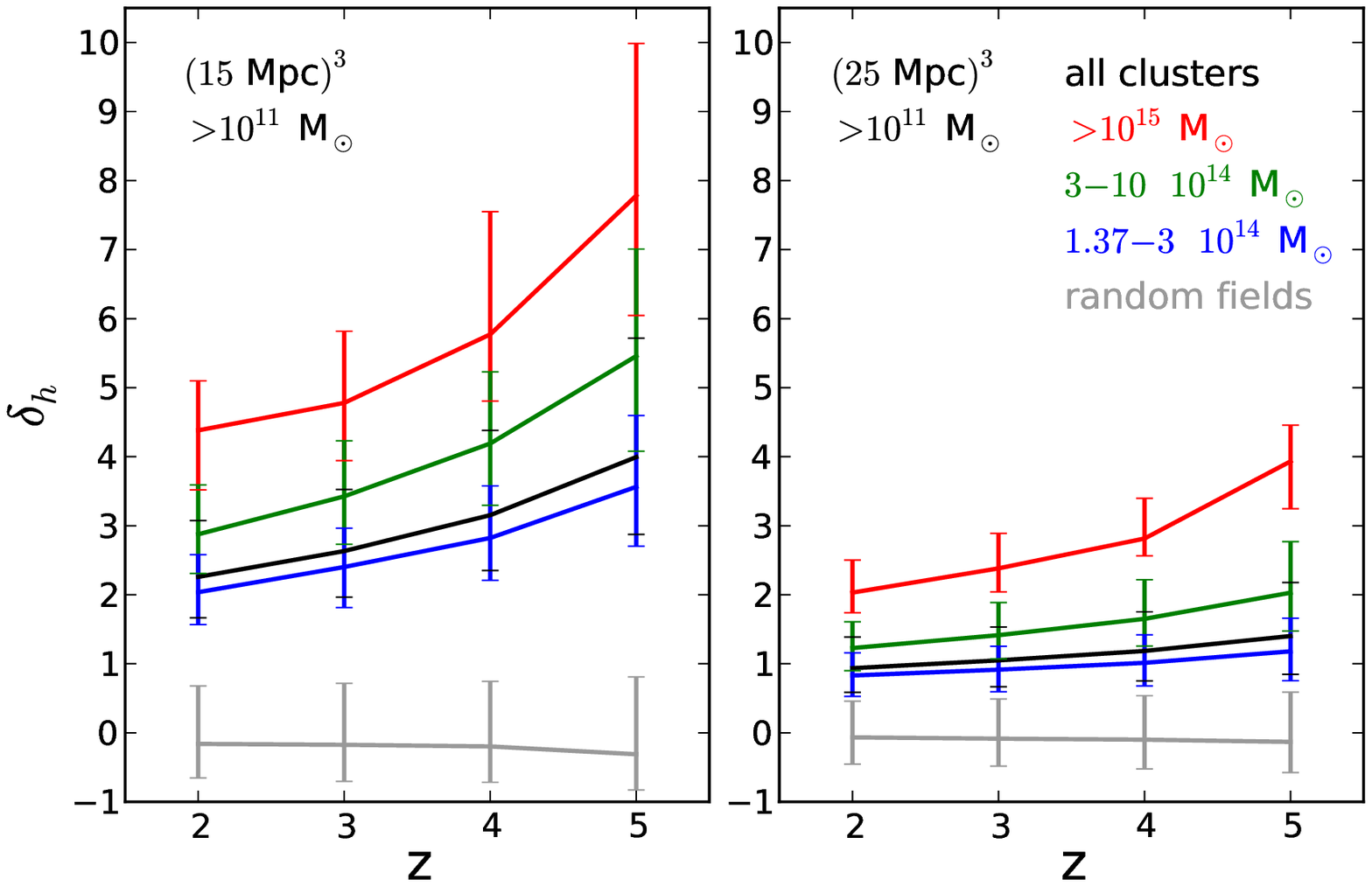}
\caption{Halo number overdensities $\delta_h$ of proto-cluster regions (black) and 10,000 random regions (gray) at redshifts 2, 3, 4, and 5 with $WMAP$1 (solid) and $WMAP$7 (dotted) cosmologies. Halos with mass $>10^{11}$ $M_{\odot}$ and windows of (15 Mpc)$^3$ (left) and (25 Mpc)$^3$ (right) comoving tophat box are used for calculating $\delta_h$. The top eight panels show the probability density distributions, $f(\delta_h)$ vs. $\delta_h$. The bottom two panels show the median $\delta_m$ and $1-\sigma$ scatter ($WMAP$1), with proto-cluster regions further binned by $M_{z=0}$. Similar to $\delta_m$, proto-clusters can be recognized by high $\delta_h$. A larger window is better to separate massive proto-clusters from lower mass ones especially at high redshifts, in the sense that the scatter and overlap in $\delta_h$ are reduced. This is because of the sizes of proto-clusters shown in Figure 2, and the fact that the $\delta_h$ statistic suffers from Poisson noise due to the discreteness.}
\end{figure}

\subsection{Halo Overdensity}
The process of galaxy formation occurs in the gravitational bound halos at the local minimums of the potential well. The overdense nature of proto-cluster regions should manifest itself not only in terms of the continuous matter distributions presented above, but also in the distribution of the individual halos and galaxies already present in that region. We test this and quantify the scatter by performing the same analysis as in the previous subsection, but now we look at the overdensity of halo number density, $\delta_h$. When counting halo number, the mass of halos needs to be specified. Here we show the case for $M > 10^{11}$ $M_{\odot}$ halos. Figure 5 shows the probability density distribution function of $\delta_h$ for random regions (gray), all (proto-)clusters (black), and (proto-)clusters binned by $M_{z=0}$ (red, green, and blue) at $z=2, 3, 4, 5$. Results with $WMAP$1 (solid) and $WMAP$7 (dotted) cosmologies are presented and they are quantitatively similar. Again, data using two different windows are plotted. Similar to Figure 4, proto-cluster regions have high $\delta_h$, and thus stand out from random fields. The most massive ones can be better identified using 25 Mpc windows. The bottom panels of Figure 5 clearly show some interesting differences from the $\delta_m$ evolution of Figure 4. Given the same minimum halo mass at different redshifts, the median values of $\delta_h$ decrease with cosmic time, even though the structures are growing. This is because halos with a certain mass at high redshift are more biased than their lower redshift ``counterparts'', and the real descendants of these high redshift halos evolve to be more massive. More biased populations are clustered stronger, and thus give higher $\delta_h$ at higher redshifts.

The choice of the limiting halo mass should be driven by observational practicalities. In order to suppress the statistical noise arising from the discrete nature of halos, one could, in principle, go to lower mass thresholds to obtain more halos. On the other hand, this may not always be possible due to the observational constraints (e.g., sensitivity of the survey).

\begin{figure}[]
\epsscale{1.2}
\plotone{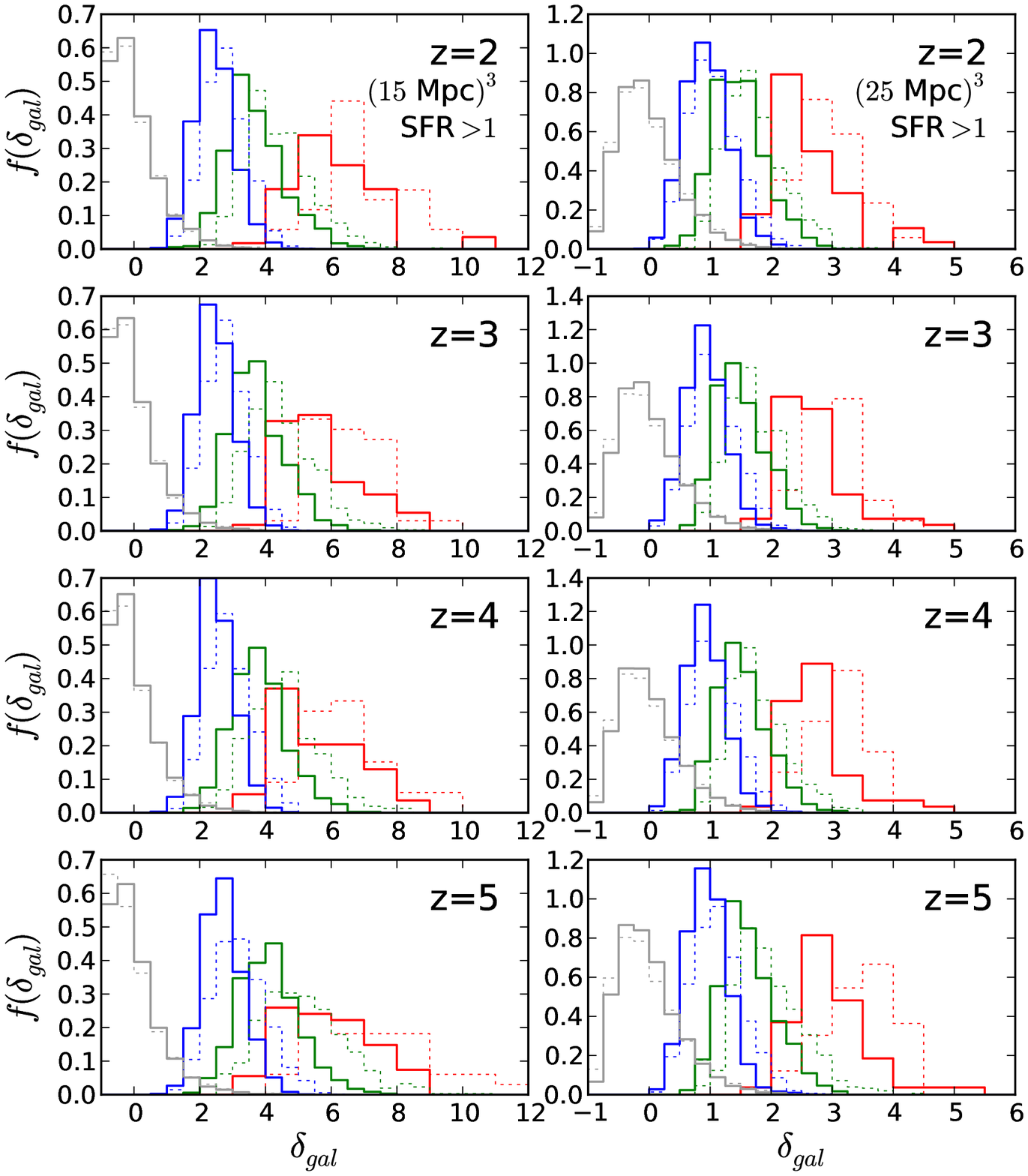}
\plotone{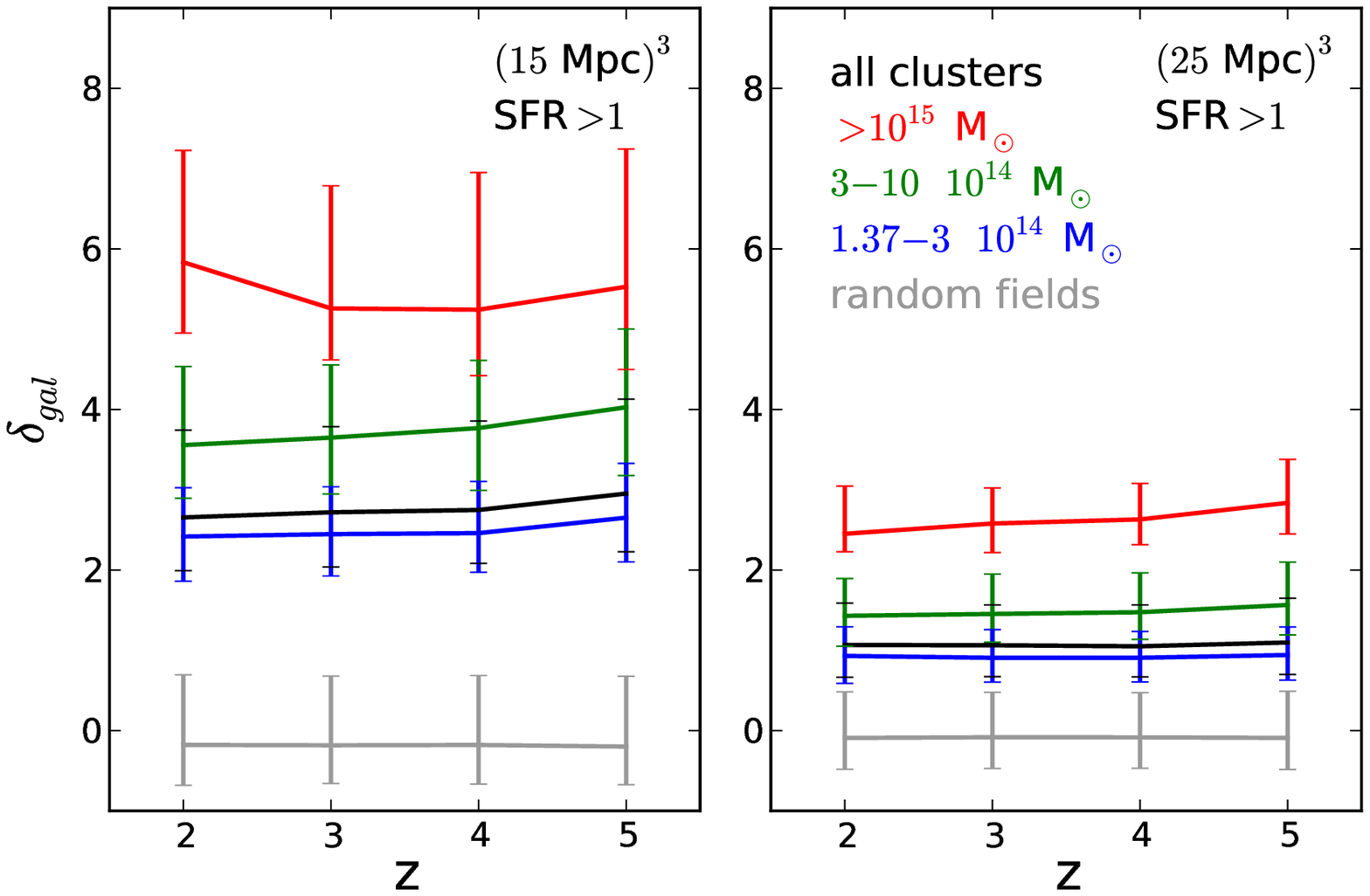}
\caption{Galaxy overdensities $\delta_{\textrm{gal}}$ of proto-cluster regions with 3 $z=0$ mass bins and 10,000 random regions (gray) at redshifts 2, 3, 4, and 5 with $WMAP$1 (solid) and $WMAP$7 (dotted) cosmologies. Galaxies with SFR $>1$ $M_{\odot}$ yr$^{-1}$ and windows of (15 Mpc)$^3$ (left) and (25 Mpc)$^3$ (right) comoving tophat box are used for calculating $\delta_{\textrm{gal}}$. The top eight panels show the probability density distributions, $f(\delta_{\textrm{gal}})$ vs. $\delta_{\textrm{gal}}$. The bottom two panels show the median $\delta_{\textrm{gal}}$ and $1-\sigma$ scatter ($WMAP$1). Proto-clusters with increasing masses can be recognized by increasing $\delta_{\textrm{gal}}$. A larger window is better to separate massive proto-clusters from lower mass ones especially at high redshifts, in the sense that the scatter and overlap in $\delta_{\textrm{gal}}$ are reduced. This is because of the sizes of proto-clusters shown in Figure 2 and the fact that the $\delta_{\textrm{gal}}$ statistic suffers from Poisson noise due to discreteness. We note that different galaxy populations can be used for calculating $\delta_{\textrm{gal}}$. In general, the distribution will peak at higher $\delta_{\textrm{gal}}$ with increasing bias of the chosen galaxy population. The scatter and noise will increase with decreasing number density of the chosen galaxy population.}
\end{figure} 

\begin{figure}[]
\epsscale{1.2}
\plotone{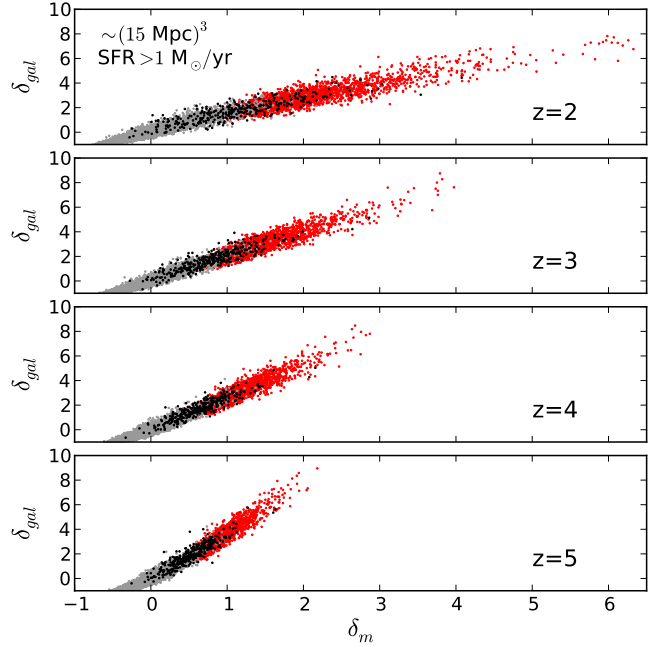}
\caption{Correlation between galaxy overdensity and the underlying mass overdensity for proto-cluster regions (red) and random regions (gray and black). Black dots are random regions that cover $>50\%$ volume (assuming a sphere with radius $R_e$) of the most nearby proto-cluster. Gray dots are random regions that are consistent with fields. Galaxies with star formation rate $>1$ $M_{\odot}$ yr$^{-1}$ are used for calculating $\delta_{\textrm{gal}}$, and windows of (15 Mpc)$^3$ comoving tophat box are applied. This correlation reflects the bias of the chosen galaxy population and it is tight at all redshifts shown here. By observing galaxy overdensities, we can infer the corresponding mass overdensities and characterize the structures.}
\end{figure}

\subsection{Galaxy Overdensity}
In the previous sections we showed that proto-clusters can be identified by mass and halo overdensity ($\delta_m$ and $\delta_h$). However, $\delta_m$ and $\delta_h$ are usually not direct observables. In order to bridge the gap between observations and theory, we therefore extract the detailed properties of proto-clusters at the level of direct observables provided by their galaxies as predicted by the SAM. We count the overdensity of galaxies in proto-clusters and random regions, with galaxies selected by different criteria such as stellar mass and SFR. Figure 6 shows the probability density distribution function of $\delta_{\textrm{gal}}$ for random regions (gray), all (proto-)clusters (black) and (proto-)clusters binned by $M_{z=0}$ (red, green, and blue). Results with $N$-body simulations and galaxy models based on $WMAP$1 (solid) and $WMAP$7 (dotted) cosmologies are presented, and they are quantitatively similar. At $z>2$, the progenitor of a ``Coma'', ``Virgo'', and ``Fornax'' type cluster is expected to have a 15/25 Mpc scale galaxy overdensity $\delta_{\textrm{gal}}\sim5.5^{+1.5}_{-0.8}$/$2.6^{+0.4}_{-0.4}$, $3.8^{+0.9}_{-0.7}$/$1.5^{+0.5}_{-0.4}$, and $2.5^{+0.6}_{-0.5}$/$0.9^{+0.3}_{-0.3}$, respectively, traced by SFR $>1$ $M_{\odot}$ yr$^{-1}$ galaxies. Remarkably, proto-clusters of different $M_{z=0}$ can be separated rather cleanly according to $\delta_{\textrm{gal}}$. This is also true when we use other selection criteria such as stellar mass (not shown here). Again, a large window size is better to identify the most massive proto-clusters. The threshold of 1 $M_{\odot}$ yr$^{-1}$ chosen here corresponds to the typical limiting SFR achieved by current surveys of Ly$\alpha$ emitters (LAEs) and LBGs (slightly more biased).

Interestingly, the time evolution of the median $\delta_{\textrm{gal}}$ are relatively flat compared to $\delta_m$ and $\delta_h$ shown previously. This is because the growth of structure and the underlying $\delta_m$ is counteracted by the decrease of galaxy bias due to selection. Given our constant SFR selection threshold at all redshifts, we are selecting a lower bias population at lower redshifts (see the galaxy bias in Table 1). This competition is also there for the case of $\delta_h$ in Figure 5. If using bright galaxies with $M_* >10^{10}$ $M_{\odot}$, a 15/25 Mpc $\delta_{\textrm{gal}}$ of $12.7^{+3.3}_{-2.8}$/$4.8^{+0.7}_{-1.2}$, $7.0^{+2.3}_{-1.5}$/$2.6^{+1.0}_{-0.7}$, and $4.4^{+1.2}_{-1.0}$/$1.6^{+0.7}_{-0.6}$ is expected at $z\sim 2$ for ``Coma'', ``Virgo'', and ``Fornax'' type proto-clusters, respectively. In this case, $\delta_{\textrm{gal}}$ increases with redshift due to the steeper evolution of galaxy bias at a fix stellar mass threshold compared to structure growth.

From the underlying mass overdensity $\delta_m$, bound units of dark matter $\delta_h$ to the final manifestation, $\delta_{\textrm{gal}}$, the overdensity--$z=0$ structure relation holds remarkably well, giving us great confidence that we can indeed use any of these criteria to find and study the early progenitors of galaxy clusters.

Next, we examine how well $\delta_{\textrm{gal}}$ can be mapped to the underlying $\delta_m$ by directly plotting $\delta_{\textrm{gal}}$ vs. $\delta_m$ (Figure 7). In this plot, red dots are regions targeted at the center of proto-clusters. Black dots are random regions that cover $> 50\%$ volume (assuming a sphere with radius $R_e$) of the most nearby proto-cluster and gray dots are random regions that are consistent with the field. The correlation between $\delta_{\textrm{gal}}$ and $\delta_m$ is very tight at all redshifts, thus it is robust to estimate $\delta_m$ from the measurement of $\delta_{\textrm{gal}}$. In order to do this, one needs to take into account the bias (defined as $b=\delta_{\textrm{gal}}/\delta_m$) of the galaxy population that was used to trace the dark matter. In Table 1 we give the bias parameters measured for the three populations of galaxies used in our analysis (i.e., samples defined as having SFR $>1$ $M_{\odot}$ yr$^{-1}$, $M_* > 10^9$, and $M_* > 10^{10}$ $M_{\odot}$). The bias was assessed at a scale of about 7.5 Mpc (i.e., half of the size of the window used to calculate the overdensities). We note that although galaxy bias in general is scale dependant, it is fairly constant at scales greater than about 1 Mpc, which corresponds to scales beyond a single halo \citep{ouchi05b}.

\begin{table}[h]
\caption{\label{tab:galaxy bias}Galaxy Bias b ($r \sim7.5$ Mpc) as a Function of Redshift and Galaxy Type}
\centering
    \begin{tabular}{ccccccccc}
        \hline\hline
         ~ & $z=2$ & $z=3$ & $z=4$ & $z=5$ \\ 
         \hline
        SFR$>1$ $M_{\odot}$ yr$^{-1}$ & 1.59 & 2.01 & 2.35 & 2.85 \\ 
        $M_{*} > 10^{9}$ $M_{\odot}$ & 1.74 & 2.24 & 2.71 & 3.38 \\
        $M_{*} > 10^{10}$ $M_{\odot}$ & 2.00 & 2.71 & 3.36 & 4.17 \\
        \hline
    \end{tabular}
\end{table}

\begin{figure}[]
\epsscale{1.2}
\plotone{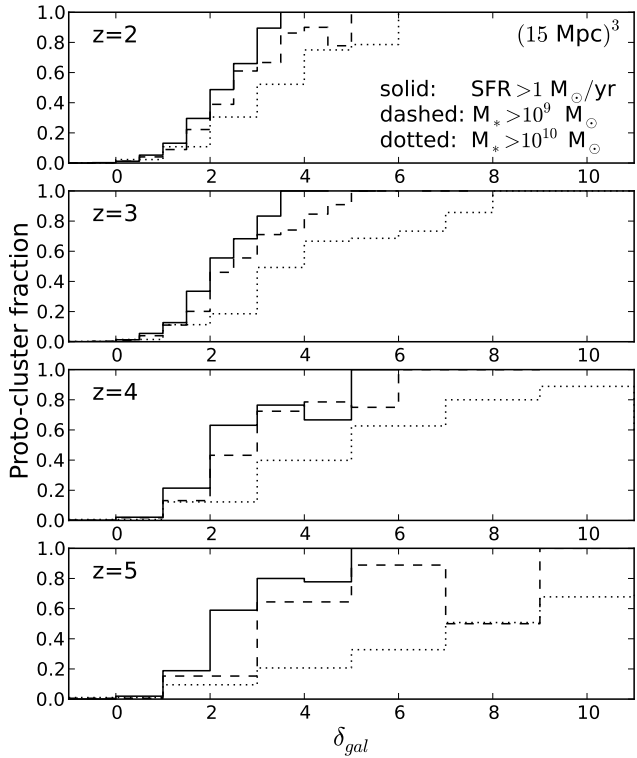}
\caption{Proto-cluster fraction as functions of galaxy overdensities calculated with different galaxy populations in windows of (15 Mpc)$^3$ comoving tophat box. This figure shows the probability for a high redshift structure to evolve into a galaxy cluster by $z=0$ given an observed $\delta_{\textrm{gal}}$. We note that the higher required $\delta_{\textrm{gal}}$ for structures to be proto-clusters using highly biased galaxies is due to the fact that highly biased populations are more clustered, making it easier to have a high $\delta_{\textrm{gal}}$.}
\end{figure}

\subsection{Identification of Proto-clusters}
As shown in Figure 7, there is a range in mass/galaxy overdensities for which the identification of a given region as a proto-cluster is ambiguous. This is because the progenitors of less massive groups sometimes show overdensities that are similar to those of small clusters, or because some (proto-)clusters are less evolved compared to other (proto-)clusters at the same epoch. In observations, this may cause field regions to be falsely classified as proto-clusters, and vice versa. To provide a tool which can be used to interpret observations at least statistically, we extract the conditional probability for a structure to finally evolve to a galaxy cluster by $z=0$ given an observed $\delta_{\textrm{gal}}$ (Figure 8). In this plot, we show the results for different galaxy selection criteria, with solid, dashed and dotted lines representing SFR $>1$ $M_{\odot}$ yr$^{-1}$, $M_*>10^9$ $M_{\odot}$ and $M_*>10^{10}$ $M_{\odot}$ galaxies, respectively. In general, the higher the observed $\delta_{\textrm{gal}}$, the more confident we are in identifying a structure as a genuine proto-cluster. When observing more biased populations, we need higher $\delta_{\textrm{gal}}$ to identify proto-clusters. This does not mean that lower bias populations are intrinsically better tracers because for a given proto-cluster, the overdensities measured from more biased populations are naturally higher. There are some concerns about the choice of tracer. First, biased populations are usually brighter and easier to detect, but the noise induced by small number statistics will propagate into the final conditional probability to identify structures. Second, a likely picture is that galaxies in dense environments may experience speed up formation and evolution, transferring less biased galaxies to be more biased. This may further reduce the strength of clustering of low bias galaxies in proto-clusters.

\begin{table*}[]
\caption{\label{tab:delta_m}$\delta_m$ Required for Achieving $50\%$ and $80\%$ Proto-cluster Fractions as a Function of Redshift and Window Size}
\centering
    \begin{tabular}{ccccccccc}
        \hline\hline
         ~ & \multicolumn{2}{c}{$z=2$} & \multicolumn{2}{c}{$z=3$} & \multicolumn{2}{c}{$z=4$} & \multicolumn{2}{c}{$z=5$} \\ 
        Window & $50\%$  & $80\%$  & $50\%$  & $80\%$  & $50\%$  & $80\%$  & $50\%$  & $80\%$  \\ \hline
        (13.4 Mpc)$^3$ & 1.35 & 2.08 & 0.93 & 1.24 & 0.77 & 1.08 & 0.59 & 0.83 \\ 
        (24.1 Mpc)$^3$ & 0.36 & 0.58 & 0.29 & 0.49 & 0.23 & 0.38 & 0.19 & 0.30 \\
        \hline
    \end{tabular}
\end{table*}

\begin{table*}[]
\caption{\label{tab:delta_h}$\delta_h$ Required for Achieving $50\%$ and $80\%$ Proto-cluster Fractions as a Function of Redshift, Window Size, and Halo Mass}
\centering
    \begin{tabular}{cccccccccc}
        \hline\hline
        ~ & ~ &  \multicolumn{2}{c}{$z=2$} & \multicolumn{2}{c}{$z=3$} & \multicolumn{2}{c}{$z=4$} & \multicolumn{2}{c}{$z=5$} \\ 
        Halos & Window  & $50\%$  & $80\%$  & $50\%$  & $80\%$  & $50\%$  & $80\%$  & $50\%$  & $80\%$  \\ \hline
        \multirow{2}{*}{$M_h > 5\times10^{10}$ $M_{\odot}$ } & (15 Mpc)$^3$ & 1.83 & 2.11 & 1.76 & 2.46 & 1.72 & 2.58 & 2.29 & 3.47 \\
        & (25 Mpc)$^3$ & 0.51 & 0.89 & 0.56 & 0.92 & 0.59 & 0.97 & 0.66 & 1.10 \\ 
        \hline
        \multirow{2}{*}{$M_h > 10^{11}$ $M_{\odot}$ } & (15 Mpc)$^3$ & 1.91 & 2.74 & 2.24 & 2.90 & 2.13 & 3.24 & 2.79 & 4.09 \\
        & (25 Mpc)$^3$ & 0.61 & 0.99 & 0.64 & 1.06 & 0.66 & 1.22 & 0.75 & 1.25 \\ 
        \hline
        \multirow{2}{*}{$M_h > 10^{12}$ $M_{\odot}$ } & (15 Mpc)$^3$ & 3.80 & 5.44 & 4.62 & 6.50 & 6.19 & 10.44 & 8.85 & $\gtrsim 25$ \\ 
        & (25 Mpc)$^3$ & 1.01 & 1.92 & 1.02 & 1.88 & 1.19 & 2.44  & 2.22 & 4.36 \\ 
        \hline
    \end{tabular}
\end{table*}

\begin{table*}[]
\caption{\label{tab:delta_gal}$\delta_{\textrm{gal}}$ Required for Achieving $50\%$ and $80\%$ Proto-cluster Fractions as a Function of Redshift, Window Size, and Galaxy Type}
\centering
    \begin{tabular}{cccccccccc}
        \hline\hline
        ~ & ~ & \multicolumn{2}{c}{$z=2$} & \multicolumn{2}{c}{$z=3$} & \multicolumn{2}{c}{$z=4$} & \multicolumn{2}{c}{$z=5$} \\ 
        Galaxies & Window & $50\%$  & $80\%$  & $50\%$  & $80\%$  & $50\%$  & $80\%$  & $50\%$  & $80\%$  \\ \hline
        \multirow{2}{*}{SFR$>1$ $M_{\odot}$ yr$^{-1}$ } & (15 Mpc)$^3$ & 2.17 & 2.88 & 2.08 & 3.14 & 2.06 & 2.83 & 1.84 & 3.04 \\
        & (25 Mpc)$^3$ & 0.62 & 1.06 & 0.64 & 1.10 & 0.62 & 0.99 & 0.60 & 1.02 \\
        \hline
        \multirow{2}{*}{$M_*>10^{9}$ $M_{\odot}$ } & (15 Mpc)$^3$ & 2.41 & 3.39 & 2.24 & 3.52 & 2.49 & 3.82 & 3.02 & 4.54 \\
        & (25 Mpc)$^3$ & 0.65 & 1.24 & 0.64 & 1.17 & 0.72 & 1.18 & 0.73 & 1.29 \\
        \hline
        \multirow{2}{*}{$M_*>10^{10}$ $M_{\odot}$ } & (15 Mpc)$^3$ & 2.87 & 4.97 & 3.43 & 4.81 & 4.29 & 6.98 & 7.38 & 10.90 \\
        & (25 Mpc)$^3$ & 0.85 & 1.42 & 0.86 & 1.49 & 0.98 & 1.85 & 1.21 & 2.69  \\
        \hline
    \end{tabular}
\end{table*}

Tables 2, 3, and 4 further list the $\delta_m$, $\delta_h$ and $\delta_{\textrm{gal}}$ required to identify a structure as a proto-cluster with $50\%$ and $80\%$ confidence for various redshifts, windows, halo, and galaxy populations. The corresponding galaxy bias is listed in Table 1.

\begin{figure}[]
\epsscale{1.2}
\plotone{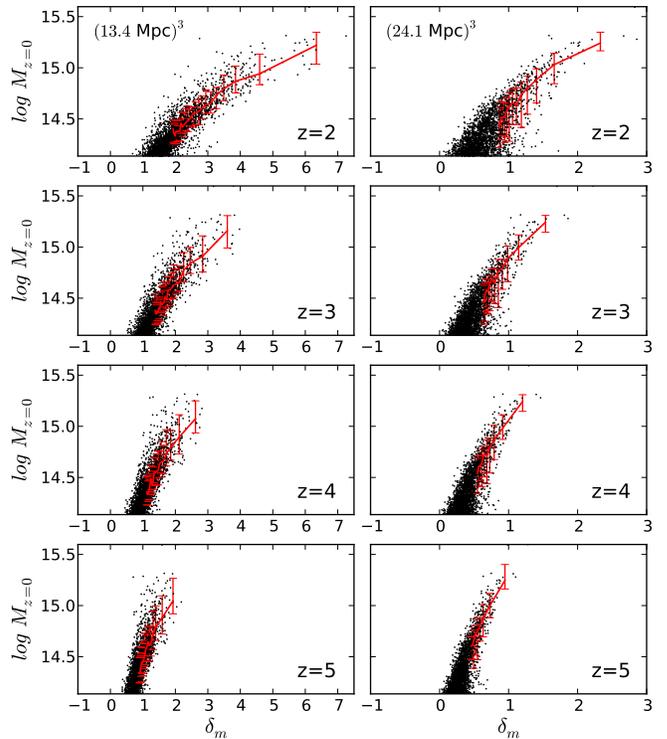}
\caption{Correlation between mass overdensity $\delta_m$ at redshifts 2, 3, 4, and 5 and the $z=0$ descendant cluster mass, $M_{z=0}$, calculated with (13.4 Mpc)$^3$ (left) and (24.1 Mpc)$^3$ (right) comoving tophat box windows. The lines and error bars indicate the median and $1-\sigma$ scatter binned by $\delta_m$. This correlation can be used to estimate the mass of proto-clusters based on their large-scale mass overdensity.
}
\end{figure}

\begin{figure}[]
\epsscale{1.2}
\plotone{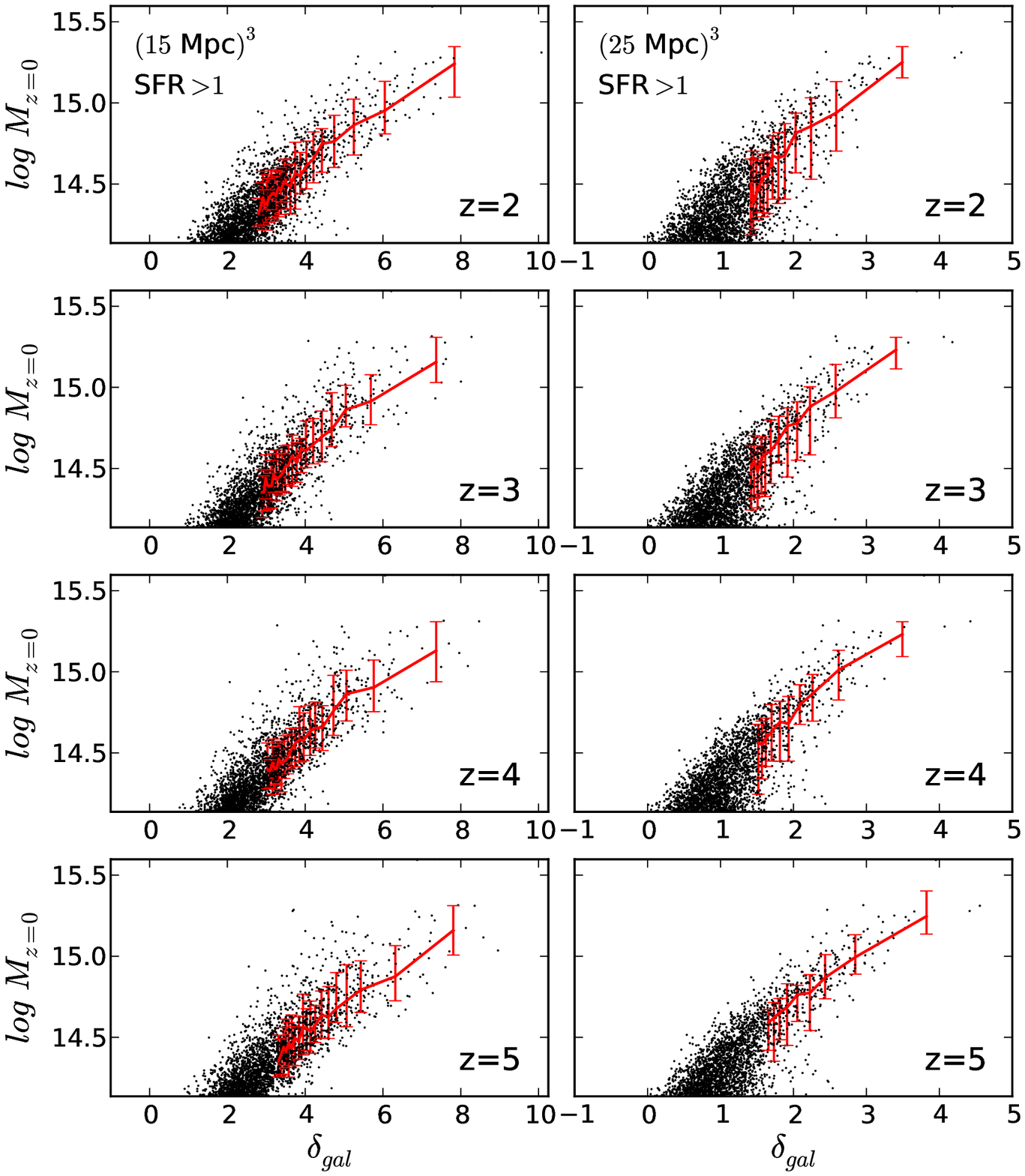}
\caption{Correlation between galaxy overdensity $\delta_{\textrm{gal}}$ at redshifts 2, 3, 4, and 5 and the $z=0$ descendant cluster mass, $M_{z=0}$, calculated with (15 Mpc)$^3$ (left) and (25 Mpc)$^3$ (right) comoving tophat box windows. Galaxies with star formation rate $>1$ $M_{\odot}$ yr$^{-1}$ are used for calculating $\delta_{\textrm{gal}}$. The lines and error bars indicate the median and $1-\sigma$ scatter binned by $\delta_{\textrm{gal}}$. This correlation can be used to estimate the mass of proto-clusters based on their observed large-scale overdensity of galaxies.}
\end{figure}

\subsection{Estimating the Present-day Masses of Proto-clusters}
 
In order to study the evolution of proto-clusters into clusters, it is extremely important that we compare structures at different redshift by statistically linking structures having a similar $M_{z=0}$. We thus need to derive reliable methods for estimating the total $z=0$ cluster mass based on the main proto-cluster observables such as their sizes and overdensities. Here we will address this problem in two ways. First, we will derive the empirical relation between overdensity and $z=0$ cluster mass based on our simulations. However, because the simulations are not necessarily representative of the true universe, we will also use them simply to test the general concept of estimating the $z=0$ descendant mass of proto-clusters based on methods suggested in literature.

The first method is to use the correlation between overdensity (of mass, halos, and galaxies) and the total $M_{z=0}$ directly. Figure 9 and 10 show the $M_{z=0}$ of our MR sample as functions of $\delta_m$ and $\delta_{\textrm{gal}}$ respectively. Again, data using two different windows and at $z=2, 3, 4, 5$ are plotted. The lines and error bars indicate the median and $1-\sigma$ scatter binned by $\delta_m$ or $\delta_{\textrm{gal}}$. As we expect, large-scale overdensities defined in fixed volume windows increase with the $M_{z=0}$. This allows us to estimate $M_{z=0}$ without further assuming a volume which contains $M_{z=0}$. A large window size, again, provides better measurements of $M_{z=0}$ for the most massive proto-clusters. In general, the errors that arise due to the intrinsic scatter are about 0.2 dex in mass for both $\delta_m$ and $\delta_{\textrm{gal}}$ (slightly larger).

Second, we perform an analysis motivated by the widely used analytic formula first applied to a $z=3.09$ proto-cluster with estimated $M_{z=0}$ of $\sim 10^{15}$ $M_{\odot}$ by \cite{steidel98, steidel00}:
\begin{equation}
M_{z=0}=(1+\delta_m) \langle\rho\rangle V,
\end{equation}
where $V$ is the true volume containing all of the mass which will be bound and virialized by $z=0$, and $\delta_m$ and $\langle\rho\rangle$ are the mass overdensity in $V$ and the average density, respectively. The $\delta_m$ is inferred from the observed galaxy overdensity, $\delta_{\textrm{gal}}$, and the linear galaxy bias, b ($\delta_{\textrm{gal}}=b \delta_m$). In principle, complexities introduced by peculiar velocities need to be considered \citep[e.g., see ][]{steidel98}.\footnote{One can perform a correction based on the Zel'dovich approximation and assumptions of spherical symmetry and dynamical stage of the collapsing structures as in Steidel et al. (1998; see also Bardeen et al. 1986; Steidel et al. 2005).} We note that all our numerical results of overdensity and volume presented in this paper refer to the true geometric positions, not considering redshift-space distortions.

\begin{figure}[t]
\epsscale{1.2}
\plotone{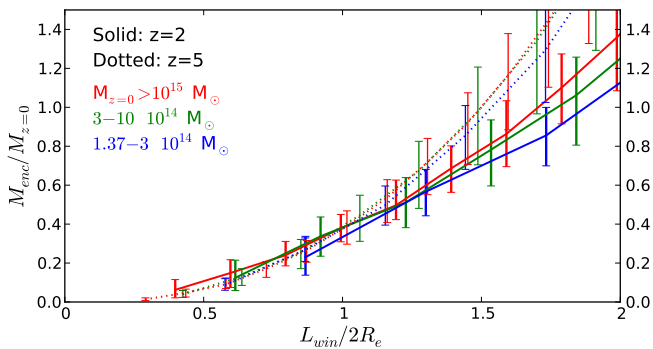}
\caption{Enclosed mass $M_{\textrm{enc}}$ within a box with a length of $L_{\textrm{win}}$ as a function of $L_{\textrm{win}}$. $M_{\textrm{enc}}$ and $L_{\textrm{win}}$ are scaled with $z=0$ descendant cluster mass and the effective diameter respectively for each (proto-)cluster. Independent of cluster mass and redshift, a proto-cluster has about $40\%$ mass inside a central cubic region of $2R_e$. This provides a universal correction factor when observationally estimating the mass of proto-clusters using an inferred density and volume.}
\end{figure} 

\begin{figure}[]
\epsscale{1.2}
\plotone{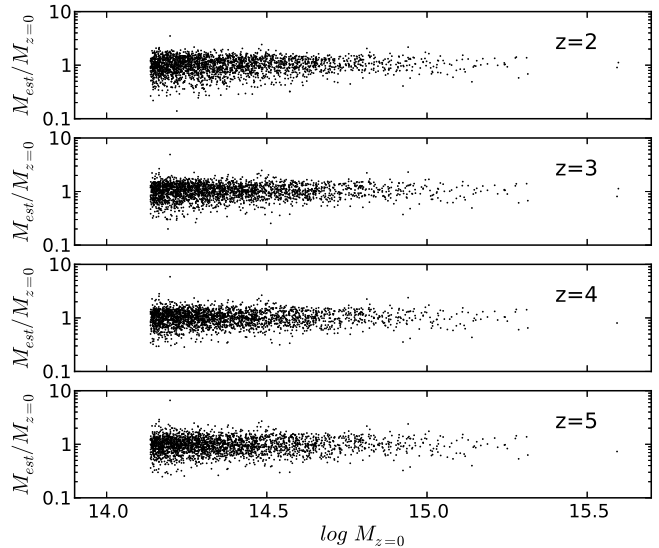}
\caption{Proto-cluster mass estimated using our proposed observational formula (Equation (6)) vs. the true $z=0$ cluster mass in the simulation. The errors given by this method are fairly low and the results show no significant bias for different intrinsic cluster masses and at different redshifts.}
\end{figure}

One of the main uncertainties in this mass estimation (Equation (5)) is the volume, which should be large enough to cover the entire structure. Also, it needs to be shown that the overdensity is large enough for the structure to collapse by $z=0$. This is usually done by translating $\delta_{m}$ (inferred from the observed $\delta_{\textrm{gal}}$ through the bias parameter), which is assumed to be described by the spherical collapse model, to the linear regime overdensity \citep[e.g., Equation (18) in][]{mo96, carroll92}, and then comparing its growth as a function of redshift with the spherical collapse threshold $\delta_{c}=1.69$ \citep[e.g.,][]{peacock99}. However, these assumptions may introduce systematic and random errors in the mass measurement for any given proto-cluster.

In order to circumvent this series of assumptions and simplifications, we propose a modification based on the insights drawn from simulations. As we proposed in Section 3.2 the effective radius, $R_e$ can serve as a characteristic size for proto-clusters, as about $65\%$ of the mass in bound halos at the concerned redshift is distributed inside $R_e$. From Equation (5), we then have:
\begin{equation}
M_{z=0} \simeq M_{\textrm{est}} \equiv C_e (1+\delta_{m, e}) \langle\rho\rangle V_e,
\end{equation}
where $V_e \equiv (2 R_e)^3$ is the effective volume and $\delta_{m, e}$ is the mass overdensity in $V_e$. $C_e$ is a correction factor that relates the mass found within the effective volume to the total mass of the cluster. It is important to note that our definition of $R_e$ is based only on the mass in bound halos found in the simulation. In reality, a significant fraction of the mass in the proto-cluster region will be in the form of smaller, unresolved halos as well as uncollapsed dark matter. In order to check the completeness of mass when applying a characteristic volume defined by the effective radius, we plot the enclosed mass $M_{\textrm{enc}}$ in cubic regions centered at the proto-cluster as a function of box size $L_{\textrm{win}}$ in Figure 11. What we found is that independent of cluster mass and redshift, a proto-cluster has about $40\%$ of mass (both bound in halos and unbound in between halos) inside a cubic region of 2$R_e$. This provides a universal correction factor when estimating the proto-cluster mass using Equation (6) of $C_e \sim 2.5$. 

Figure 12 shows the total mass estimated by our proposed $M_{\textrm{est}}$ (Equation (6)) across the mass and redshift range of interest. As we see $M_{\textrm{est}}$ works fairly well to reproduce the intrinsic mass $M_{z=0}$ . The errors are in general less than a factor of 2 and the results show no significant bias for different intrinsic cluster masses and at different redshifts.

Practically, $M_{\textrm{est}}$ can be observationally obtained using derived $\delta_m$ and a volume corresponding to $R_e$. By definition (Equation (4)), $R_e$ is based on spatial distribution of halo mass, which can be estimated from the observed spatial distribution of galaxies and their inferred halo masses, or by mapping overdensity profiles of proto-cluster galaxies (see Figure 3).

\subsection{Effects of Redshift Measurement Uncertainty}

In some of the real observations traced by narrow-band-selected emission line galaxies or samples with photometric redshift, galaxy overdensity is basically measured in excess of surface density, $\delta_{\textrm{gal}}=(\Sigma-\bar{\Sigma})/\bar{\Sigma}$. In these cases, the full width of the redshift range, $\Delta z$, for the narrow-band filter or the photometric redshift uncertainty equivalently sets a window with a depth that is often larger in the radial dimension than that assumed by our 15 or 25 Mpc cubic windows. As $\Delta z$ increases, the projection of the low density proto-cluster outskirts and physically unassociated interlopers diminishes the significance of proto-cluster overdensity, while the projection of overlapping structures can also spuriously boost the observed overdensity. We demonstrate and quantify this effect by adjusting the depth of the applied windows to the comoving length $d_c$ set by given $\Delta z$. Figure 13 shows the $\delta_{\textrm{gal}}$ as a function of $\Delta z$ for proto-clusters and 10,000 random regions at $z=3$ traced by galaxies with $\textrm{SFR} >1$ $M_{\odot}$ yr$^{-1}$. The effectively larger windows with increasing $\Delta z$ smooth the density field, decrease the scatter within each bin, and largely diminish the proto-cluster overdensity. For example, for star-forming Ly$\alpha$ emitting galaxies at $z=3$ selected with a typical narrow-band filter (FWHM $\sim 60$ $\text{\AA}$, corresponding to a $\Delta z$ $\sim 0.05$ and $d_c \sim 50$ Mpc), the surface density of proto-clusters ($1+\delta_{\textrm{gal}}$) drops $\sim$ 40\% from the values calculated with a 15 Mpc cubic window. At a $\Delta z$ $\gtrsim 0.1$, it becomes difficult to distinguish proto-clusters from random fields, except for the most overdense systems. On the other hand, if accurate spectroscopic redshifts are obtained, one might still need to correct for the redshift-space distortion due to the intrinsic velocity dispersion of proto-cluster galaxies \citep[e.g.,][]{steidel98} if the desired window depth to calculate $\delta_{\textrm{gal}}$ is relatively small. The line of sight velocity dispersion of our ``Coma-'', ``Virgo-'', and ``Fornax-'' type sample at $z\sim3$ are at the level of $400\pm60$, $330\pm70$, and $250\pm60$ km s$^{-1}$, respectively, and they are $\sim 20\%$--$40\%$ higher at $z=2$.

\begin{figure}[]
\epsscale{1.2}
\plotone{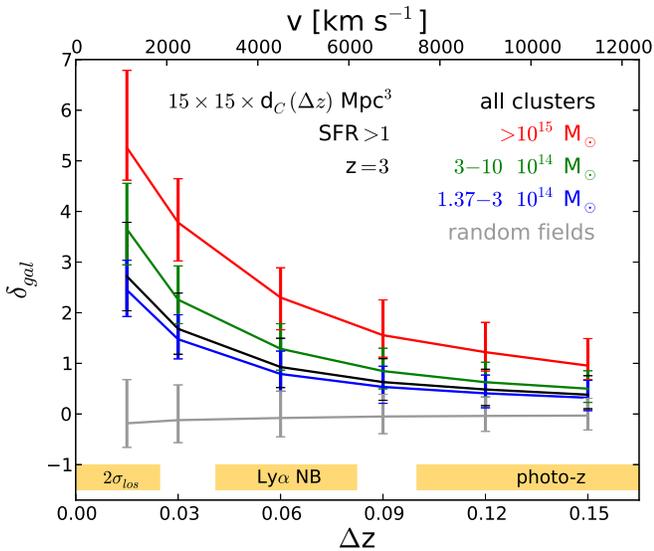}
\caption{Median and $1-\sigma$ scatter galaxy overdensities $\delta_{\textrm{gal}}$ for proto-clusters and random regions as function of redshift uncertainty $\Delta z$ (full width) at $z=3$. Windows with $15\times15$ Mpc$^2$ on the sky times a comoving radial depth corresponding to $\Delta z$ are used. The depth for the smallest $\Delta z$ data points is set to 15 Mpc, thereby recovering the same $\delta_{\textrm{gal}}$ as shown in Figure 6. Galaxies with $\textrm{SFR} >1$ $M_{\odot}$ yr$^{-1}$ (bias $b \sim2$ at $z=3$) are used to calculate $\delta_{\textrm{gal}}$. As the $\Delta z$ increases, the scatter for each population gradually decreases due to smoothing, while the mean $\delta_{\textrm{gal}}$ for clustered populations goes down rapidly and eventually becomes indistinguishable from the random fields. The shaded bars indicate the approximate redshift uncertainties allowed by typical spectroscopic, narrow-band, and color-selection techniques.}
\end{figure}

In Figure 13, we use shaded bars to mark the different regimes that can be probed by different selection techniques: (1) a spectroscopic survey in which the redshift uncertainty is set by the velocity distribution of the proto-cluster (full width of 2$\sigma_{\textrm{los}}$ and assuming $\sigma_{\textrm{los}}$ of up to 460 km s$^{-1}$), (2) a narrow-band Ly$\alpha$ survey using a typical narrow-band filter of FWHM $50$--$100$ $\text{\AA}$, and (3) a photometric redshift survey for which $\Delta z$ $\gtrsim0.1$. It is worth noting that although they will be highly incomplete, techniques with large $\Delta z$ are still valuable for selecting candidates of proto-cluster. These candidates will then likely be the progenitors of more massive clusters, as shown in Figure 13. Such structures will be interesting ``by-products'' of upcoming dark energy galaxy surveys.

\section{Discussion}

The search for proto-clusters and the characterization of their main properties offer great challenges and opportunities for the study of cluster formation. Complementary to ongoing observational efforts, our results demonstrate that invaluable insights and quantitative descriptions can also be gained from simulations. For example, in this paper we have given, for the first time, detailed predictions based on $\sim3000$ simulated galaxy clusters, while only $\sim20$ proto-clusters (or candidates thereof) are known from observations.
What have we learned from these simulations predictions? As shown in Figure 2, at $z>2$ most cluster progenitors are not yet virialized at the mass scale of present-day clusters (defined as having a mass $\gtrsim 10^{14}$ $M_{\odot}$). This is consistent with the absence of extended X-ray emission in observations beyond $z\sim2$ to date. However, according to Figure 2, the rarest progenitors of the most massive ``Coma-type'' clusters formed as early as $z \sim 2.3$, indicating the possibility of X-ray emitting ICMs yet to be discovered at $z>2$. This is also suggested by the hydro-dynamical simulations performed by \cite{saro09}. However, despite the fact that luminous or detectable levels of X-ray emission will be absent for most forming clusters at such high redshifts, we have clearly demonstrated that one can, in principle, already identify the progenitors of massive clusters out to much higher redshifts by simply focusing on their overdensities of galaxies. Once identified, it will be much more efficient to target these structures for faint, diffuse X-ray emission rather than to perform deep blind searches with current X-ray missions.

\begin{table*}[ht]
\caption{\label{tab:protoclusters}An Overview of Protocluster Candidates Select from the Literature} 
\footnotesize
\begin{center}
\begin{tabular}[t]{lclcccccc}
\hline
\hline	\multicolumn{1}{c}{Object} & \multicolumn{1}{c}{$z$} &\multicolumn{1}{c}{Sample$^a$} &  \multicolumn{1}{c}{Window Size$^b$} &  \multicolumn{1}{c}{$\Delta z^c$} & \multicolumn{1}{c}{$\delta_{\textrm{gal}}^d$} & \multicolumn{1}{c}{$\sigma_v^e$} & \multicolumn{1}{c}{$M^f$} &  \multicolumn{1}{c}{References$^\dagger$} \\
   & & &  (arcmin$^2$) & & & (km s$^{-1}$) & ($10^{14}$ M$_\odot$) & \\
\hline
PKS 1138--262        &   2.16     &  \lya                     &  $7\times7$     &	0.053	&   $3\pm2$      &  $900\pm240$      & 3--4      &   1,2,3,4,5,6   \\[1ex]
        &       &  H$\alpha$                     &  $7\times7$     &	0.041	&   $\dots$     &  $\dots$     & $\dots$      &   7   \\[1ex]

HS1700--FLD         &   2.30    &  BX                      &  $8\times8$     &		0.030	&  $6.9^{+2.1}_{-2.1}$       & $\dots$      & 14       &   8  \\[1ex]

4C 10.48         &   2.35    &  H$\alpha$                      &  $2.5\times2.5$     	&	0.046	&  $11^{+2}_{-2}$       & $\dots$      &    $\dots$    &  9   \\[1ex]

J2143--4423         &   2.38  &  \lya       &  $44\times44$     &	0.044	&  $5.8^{+2.5}_{-2.5}$       & $\dots$      &    $\dots$    &   10  \\[1ex]

4C 23.56         &   2.48    &  H$\alpha$                      &  $7\times4$     &	0.035	&  $4.3^{+5.3}_{-2.6}$       & $\dots$      &    $\dots$    &  11   \\[1ex]

USS 1558--003         &   2.53    &  H$\alpha$                      &  $7\times4$     &	0.041	&  $\dots$       & $\dots$      &    $\dots$    &   12  \\[1ex]

LABd05      & 2.7        &  \lya    &    $28\times11$                  &	0.165	&    $\sim2$     &  $\dots$       & $\dots$      & 13        \\[1ex]

HS1549      & 2.85        &  LBG    &    $\dots$                  & 	0.060	&   $\sim5$     &  $\dots$    & $\dots$      & 14        \\[1ex]

MRC 0052--241      & 2.86        &  \lya    &    $7\times7$                  &	0.054	&    $2.0^{+0.5}_{-0.4}$     &  $980\pm120$       & 3--4      &  6,15    \\[1ex]

MRC 0943--242      & 2.92        &  \lya    &    $7\times7$                  &	0.056	&    $2.2^{+0.9}_{-0.7}$      & $715\pm105$       & 4--5     &  6,15    \\[1ex]

SSA22--FLD           &   3.09     &  LBG     &   $11.5\times9$                    &  0.034	& $3.6^{+1.4}_{-1.2}$      &    $\dots$    & 10--14     &   16\\
                    &             &  \lya    &   $9\times9$                     &	0.066	&  $5\pm2$       &  $\dots$      & $\dots$                    &     17,18,19 \\[1ex]
MRC 0316--257        &   3.13      &  \lya    &   $7\times7$                     &	0.049	&   $2.3^{+0.5}_{-0.4}$       &  $640\pm195$       &3--5       &   6,15,20\\[1ex]

TN J2009--3040       &   3.16      &  \lya    &      $7\times7$                 &	0.049	&      $0.7^{+0.8}_{-0.6}$    &   $515\pm90$     & $\dots$ &  6,15     \\[1ex]

TN J1338--1942       &   4.11      &  \lya    &    $7\times7$ ($\times2$)                    &  0.049	&  $3.7^{+1.0}_{-0.8}$     &   $265\pm65$     &6--9      &   6,15,21\\
                    &             &  LBG     &     $3.4\times3.4$                   &   	$\sim 0.6$	& $1.5^{+0.3}_{-0.3}$     &  $\dots$      & $\dots$      &   6,22,23,24\\[1ex]

6C 0140+326              &   4.41      &  \lya    &    $10\times10$                    & $\sim 0.04$	&   $8^{+5}_{-5}$     &   $\dots$     &  0.8--2.9  &     25\\[1ex]

SDF              &   4.86      &  \lya    &    $10\times10$                    &	0.060	&    $2.0^{+1.0}_{-2.0}$     &   $\dots$     &  $>3$  &     26\\[1ex]

TN J0924--2201       &   5.19      &  \lya    &     $7\times7$                    & 	0.073	&   $1.5^{+1.6}_{-1.0}$     &  $305\pm110$      & 4--9    &   6,15,27\\
                    &             &  LBG     &      $3.4\times3.4$                   &    $\sim 0.7$		& $1.0\pm0.5$     &   $\dots$     &   $\dots$ &    28\\[1ex]

COSMOS AzTEC03               &   5.30       &  SMG   &   $1\times1$                     	&  $\dots$      &  $\dots$   & $\dots$   &  $\dots$  &   29  \\[1ex]

SXDF-Object `A'$^g$                &   5.70       &  \lya    &   $6\times6$                     &	0.099	& $3.3^{+0.9}_{-0.9}$      &  $\sim180$      &  1--3  &     30\\[1ex]

SDF              &   6.01      &  LBG    &    $6\times6$                    &  $\sim 0.05$  	&	$16\pm7$     &   $647\pm124$    &  2--4  &     31\\[1ex]

CFHQSJ2329--0301                &   6.43       &  LBG    &   $34\times27$                     & 	$\sim 1.0$	& $\sim 6$      &   $\dots$    &  $\dots$  &  32  \\[1ex]

See also                &          &    &              	&    &     & 	&   &  33,34,35,36,37  \\[1ex]

\hline
\end{tabular}
\end{center}
\begin{scriptsize}
\textbf{Notes.}\\
$^{a}${Method of sample selection: (\lya) narrowband \lya, (H$\alpha$) narrowband H$\alpha$, (LBG) Lyman break technique, (BX) the `BX' criteria of Adelberger et al. (2005), (SMG) sub-millimeter galaxies.}\\
$^{b}${Approximate field size or the size of the structure used to calculate overdensity.}\\
$^{c}${Full width redshift uncertainty associated with the $\delta_{\textrm{gal}}$ quoted.}\\
$^{d}${Amplitude of the galaxy overdensity in the references. Except for HS1700--FLD, SSA22--FLD, and SDF ($z=6.01$) where ample spectroscopic information was available, $\delta_{\textrm{gal}}$ refers to the projected surface overdensity $(\Sigma-\bar{\Sigma})/\bar{\Sigma}$.}\\
$^{e}${Velocity dispersion (where available).}\\
$^{f}${Inferred mass of the overdensity in units of $10^{14}$ M$_\odot$.}\\
$^{g}${Only the richest of the two $z=5.7$ overdensities discovered in this field is listed.}\\
\textbf{References.} $^\dagger${(1) Kurk et al. 2000; (2) Pentericci et al. 2000; (3) Pentericci et al. 2002; (4) Kurk et al. 2004a; (5) Kurk et al. 2004b; (6) Venemans et al. 2007; (7) Koyama et al. 2013a; (8) Steidel et al. 2005; (9) Hatch et al. 2011; (10) Palunas et al. 2004; (11) Tanaka et al. 2011; (12) Hayashi et al. 2012; (13) Prescott et al. 2008; (14) Mostardi et al. 2013; (15) Venemans et al. 2005a; (16) Steidel et al. 1998; (17) Matsuda et al. 2005; (18) Steidel et al. 2000; (19) Yamada et al. 2012; (20) Venemans et al. 2005b; (21) Venemans et al. 2002; (22) Miley et al. 2004; (23) Zirm et al. 2005; (24) Overzier et al. 2008; (25) Kuiper et al. 2011; (26) Shimasaku et al. 2003; (27) Venemans et al. 2004; (28) Overzier et al. 2006b; (29) Capak et al. 2011; (30) Ouchi et al. 2005; (31) Toshikawa et al. 2012; (32) Utsumi et al. 2010; (33) Papovich et al. 2012; (34) Chiaberge et al. 2010; (35) Spitler et al. (2012; (36) Matsuda et al. 2009; (37) Trenti et al. 2012.}\\
 
\end{scriptsize}
\end{table*}

In order to initiate a more systematic study of proto-clusters at the present moment, we have compiled an overview of known structures selected from the literature. The results are shown in Table 5. To our knowledge, about 20 good proto-cluster candidates have been found to date. These structures have redshifts in the range $2\lesssim z \lesssim 6$ and galaxy overdensities in the range $\approx 1\textrm{--}16$ on scales from a few to a few tens of arcminutes. Furthermore, because these proto-clusters were found using a wide range of tracer galaxies, the sample is rather inhomogeneous and somewhat difficult to compare with any single set of simulation predictions. However, the main observable properties of these structures (i.e., the overdensity and the projected size) generally agree with the typical properties of the proto-clusters in the simulations described in this paper.

About half of the structures were found around high redshift radio galaxies. We compare our results with the systematic study of proto-clusters associated with powerful radio galaxies \citep[see][and Table 5]{venemans07}. For example, four structures at $z\sim 3$ (MRC 0052--241, MRC 0943--242, MRC 0316--257, and TN J2009--3040) showed surface overdensities of LAEs of $2.0^{+0.5}_{-0.4}$, $2.2^{+0.9}_{-0.7}$, $2.3^{+0.5}_{-0.4}$, and $0.7^{+0.8}_{-0.6}$, respectively. The \citet{venemans07} study was performed with a narrow-band filter ($59\textrm{--}68$ $\text{\AA}$, FWHM) over a field of view of $7 \times 7$ arcmin$^2$ ($\sim13 \times 13$ cMpc$^2$). We can compare this study with our predictions shown in Figure 13 taking $\Delta z$ $\sim 0.05$ corresponding to the redshift range probed by the narrow-band filter. Based on Figure 13, a ``Coma'', ``Virgo'', and ``Fornax'' type proto-cluster is then expected to have a $\delta_{\textrm{gal}}$ of $2.8^{+0.6}_{-0.7}$, $1.6^{+0.6}_{-0.5}$, and $1.0^{+0.5}_{-0.4}$ given this approximate set-up. Therefore, MRC 0052--241, MRC 0943--242, and MRC 0316--257 are likely to be  progenitors of a ``Virgo-type'' cluster with $M_{z=0} \lesssim 10^{15}$ $M_{\odot}$. TN J2009--3040 is likely to be the progenitor of a low-mass cluster or massive group, given its much lower overdensity. We also find that the present-day masses ($M_{z=0}$) as estimated by \cite{venemans07} are systematically lower than our results by a factor of $\sim 2$. This can be explained by a relatively high value for the galaxy bias parameter ($b=3\textrm{--}6$) that they assumed when converting between $\delta_{\textrm{gal}}$ and $\delta_m$. More recent observations have found that LAEs are likely to be less biased having $b=1.7^{+0.3}_{-0.4}$ at $z\sim 3$ \citep[e.g.,][]{gawiser07}, similar to the value of the Ly$\alpha$-like star-forming sample used in our simulations ($b \sim 2$). Therefore, our results match those obtained by \cite{venemans07} if we apply the updated bias value to their results. 

We also compare our simulations with one of the best known examples of proto-clusters at $z \sim 4$, the one associated with radio galaxy TN J1338--1942 at $z=4.11$ (see Table 5 for references). TN J1338--1942 shows a $\delta_{\textrm{gal}}$ of $\sim 4$ measured over a field of $\sim 20\times 20$ cMpc$^2$ traced by LAEs found within a narrow-band filter that has a width corresponding to $\Delta z \sim 0.05$. The approximate value for the bias parameter of $z\sim4$ LAEs is $\sim3\textrm{--}4$ \citep{kovac07, ouchi10, jose13}. If we tune our simulations predictions to this particular observational configuration (not shown here) and compare them with the observed properties of TN J1338--1942, we find that it is most likely the progenitor of a ``Virgo-type'' galaxy cluster. 
This is consistent with \cite{venemans07}, who estimated that $M_{z=0} = 6\textrm{--}9 \times 10^{14}$ $M_{\odot}$ based on simple analytical arguments (e.g., using Equation (5)). Furthermore, from Figure 2 we find that this type of proto-cluster is expected to first pass the cluster mass ``threshold'' of $10^{14}$ M$_\odot$ near $z\sim1$.


Despite the convenience to evaluate $\delta_{\textrm{gal}}$ within windows of a fixed size, it is important to study the true extension and topology of overdensities. Based on our results presented in Section 3.2 and 3.3, proto-clusters can be significantly larger than assumed in some of the literature which has focused on trying to identify only the core regions \citep[e.g.,][]{hatch11a}. Although regions exist that show large overdensities on relatively small scales, these will be more prone to projection effects or confusion with the field, and they offer less leverage in determining the true scale of the overdensity. For example, if a $z\sim 2$ overdense clump of size $\sim 1$ physical Mpc is to be the progenitor of a $10^{15}$ $M_{\odot}$ cluster, it is likely that this clump is surrounded by a significant overdensity out to $\gtrsim 20$ comoving Mpc across with an effective diameter of 2$R_e=13.0^{+3.8}_{-2.6}$ cMpc that encompasses about 40\% of the mass. If such a large-scale overdensity is not seen, then it is most likely the progenitor of a much less massive cluster or group. On the other hand, extremely large-scale structures with a scale of $\sim50\textrm{--}60$ comoving Mpc have also been found at high redshift \citep[e.g.,][]{shimasaku03, matsuda05}. Although the central parts of these structures might form massive virialized clusters by $z=0$, it is unlikely that the collapse of the entire structure will have been completed by $z=0$.


The present-day mass of the descendant cluster, $M_{z=0}$, is the main physical quantity that should be used to link proto-clusters at high redshift to clusters at low redshift. As we have shown in this paper, many properties such as the size, virialization redshift and overdensity correlate well with $M_{z=0}$. Therefore, $M_{z=0}$ can serve as the principle parameter to classify and characterize structures across the ``proto-cluster zoo'' summarized in Table 5. Galaxy formation processes and time scales are expected to systematically differ along this mass sequence. For example, cluster red sequences are expected to form earlier in more massive proto-clusters. Using the overdensity--cluster mass ($\delta_{\textrm{gal}}$--$M_{z=0}$) relation presented in Figures 9 and 10, as well as the calibrated volume--overdensity approach (Equation (6)) with a correction factor $C_e \sim 2.5$, we have shown that several, relatively safe methods exist for estimating $M_{z=0}$ with a small intrinsic scatter ($\sim 0.2$ dex, Figure 12). While the most popular method for estimating the total mass of proto-clusters is currently through the use of Equation (5), this method has a disadvantage in that it requires knowing not only the overdensity, but also the total size of the volume that will have collapsed at $z \sim 0$. Our first method presented in Figures 9 and 10 circumvents this complication by directly calibrating the relation between overdensity and $M_{z=0}$ for a fixed window size. The main disadvantage of course is that this relation is not simulations-independent. However, as simulations improve, so will the calibration. In the mean time, it should be possible to use the predictions to establish at least a relative mass scale for proto-clusters found in observations. 

It is important to note that a very large number of proto-clusters are needed before we can draw any statistically significant conclusions related to cluster formation from observations, due to the relatively large dispersion in, e.g., the sizes and overdensities found even for progenitors of clusters having the same mass at $z=0$. As we have demonstrated in Section 3.9 and Figure 13, the uncertainty in the redshift measurements largely diminishes the feasibility to distinguish proto-cluster regions from random fields due to projection and smoothing. Therefore, the ideal tracer galaxies for proto-clusters are not necessarily the most abundant galaxy population in proto-clusters. Emission line galaxies such as LAEs are good tracers provided that their spectroscopic redshifts are obtained. On the other hand, it is crucial to get a better handle on the bias of galaxy tracers since the statistical and systematic errors will directly propagate into the derived physical properties of the proto-clusters. A unique upcoming large area Ly$\alpha$ survey, the Hobby-Eberly Telescope Dark Energy Experiment (HETDEX, PI: G. Hill), may generate the first large sample of hundreds to thousands of proto-clusters at $1.9<z<3.5$. HETDEX is going to perform a blind integral-field unit  spectroscopic survey using the upgraded 9.2m Hobby-Eberly Telescope to probe a $\sim$10 Gpc$^3$ volume. Likewise, the Hyper Suprime-Cam and the Prime Focus Spectrograph on the Subaru Telescope will perform deep, large area imaging and spectroscopic surveys of LBGs and LAEs allowing the discovery of a large number of proto-clusters and associated galaxies. Meanwhile, the large volumes and precise redshift information of these surveys will largely improve the constraint on galaxy bias. Based on the framework presented in this paper, we will be able to construct the first large statistical samples of proto-clusters from such surveys. Then we will be able to estimate their $z=0$ masses and construct bins corresponding to the progenitors of different present-day mass clusters. This will finally allow us to systematically compare the properties of clusters and their galaxies at different redshifts and perform a full census of cluster evolution during the ``cosmic noon''.

\section{Summary}

In order to pave the way for large, statistical studies of galaxy cluster formation that will be possible with upcoming surveys, we have data-mined the $\Lambda$CDM MR dark matter and semi-analytic simulations to study the progenitors of 2832 galaxy clusters ($M_{z=0} > 10^{14}$ $M_{\odot}$ $h^{-1}$) and their galaxy populations across cosmic history ($0<z<5$). In this first paper, we present, for the first time, the bulk properties such as the evolution in total mass, size, and overdensity of proto-cluster regions as a function of, e.g., redshift and present-day cluster mass. Our main findings are as follows.

1. A proto-cluster can be defined as a large-scale structure which will evolve into a galaxy cluster by $z=0$, and its $z=0$ mass
($M_{z=0}$) is closely related to its main properties at all redshifts (e.g., size, dark matter and galaxy overdensity, virialization redshift). Before the cluster assembly redshift (defined as the redshift at which the proto-cluster first contains a halo of $10^{14}$ M$_\odot$), the structure is not virialized on a cluster-scale. The  basic observational features of the proto-cluster are (1) one or a few massive halos and galaxies in the core region of the overdensity, and,  more importantly; (2) a significant overdensity in mass, halos, and galaxies that extends out to very large comoving scales (many Mpc) that can already be identified in the large-scale structure as early as $z\sim5$.

2. In order to assess when proto-clusters first became clusters, we track the evolution of each cluster's most massive progenitor halo over 13 Gyr (Figure 2, left). If we follow the convention and define a galaxy cluster as a bound object with a mass that exceeds $10^{14}$ $M_{\odot}$, different (proto-)clusters first pass this threshold in the redshift range from $z=0.2$ to $z=2.3$ depending on their final mass. The typical cluster with a mass similar to that of the Virgo cluster passes this threshold at $z\sim1$. However, the most massive clusters (Coma-type clusters or more massive) reach the threshold mass as early as $z\approx2.3$, indicating that massive, and perhaps X-ray luminous, structures already exist at these early times.

3. We define an effective radius $R_e$ (Equation (4)) that encapsulates $\sim$65\% (40\%) of the mass in halos (total mass) of a proto-cluster. The proto-cluster effective diameter $2R_e$ is in the range $\sim5\textrm{--}22$ Mpc at $2<z<5$ (Figure 2, right). Progenitors of more massive clusters are larger than those of less massive clusters at all redshifts, while the progenitors regions of all clusters were larger at higher redshifts.

4. We quantify the overdensities associated with proto-clusters in terms of the dark matter, dark matter halos, and galaxies as a function of $M_{z=0}$, redshift, selection window size, and various halo and galaxy selection criteria (Figures 3--5). By comparing with random regions, we derive the conditional probability for an observed structure being a true proto-cluster given a set of observables. Our predictions based on large-scale galaxy overdensities are particularly useful, as we lack the means of directly measuring dark matter mass at high redshifts. 

5. We present two estimators for deriving $M_{z=0}$ based on  the correlation between $M_{z=0}$ and the observed galaxy overdensity (Section 3.8). We show that the mass of present-day clusters can be ``predicted'' from the observed galaxy overdensity at high redshift with an intrinsic scatter of $\sim0.2$ dex in $M_{z=0}$ (Figures 9--12). This is promising for future studies as it will allow us to study the evolution of clusters all the way from the proto-cluster phase to their present-day state, properly binned in redshift and (present-day) cluster mass.

6. Projection effects arise when the data allows one to only measure galaxy surface overdensities instead of volume overdensities. Although this has a minor effect on the significance of the overdensities measured from narrow-band-selected or spectroscopic samples, it has a major effect on our ability to identify and correctly classify proto-clusters using much cruder selections based on (broad-band) color selection (see Figure 13).

7. We present and discuss a wide range of proto-clusters (and candidates thereof) selected from the literature (see Table 5). In general, these structures with galaxy overdensities of order of a few measured over fields 10--20 Mpc (comoving) in size, are very similar to those predicted by our simulations, indicating that they may indeed represent the earliest stages of cluster formation.

8. Our work demonstrates the feasibility of extending current studies of cluster evolution at low redshift into the epoch at $z\gtrsim2$ where the clusters and their galaxies were actually forming. The first large statistical studies of such systems can be undertaken with data from upcoming surveys such as HETDEX and Subaru/Hyper Suprime-Cam that should generate very large maps of the large-scale structure in three dimensions based on LAEs and LBGs.

\begin{acknowledgements}
We thank Eduardo Ba$\rm \tilde{n}$ados, Peter Behroozi, Chi-Ting Chiang, Tiago A. Costa, Steven Finkelstein, Nina Hatch, Shardha Jogee, Lindsay King, Milos Milosavljevic, Masami Ouchi, Casey Papovich, Emma Rigby, Huub R$\rm \ddot{o}$ttgering, Lee Spitler, Bram Venemans and Risa Wechsler for helpful discussions. Support for this work was provided by NASA through an award issued by JPL/Caltech. The Millennium Simulation databases used in this paper and the web application providing online access to them were constructed as part of the activities of the German Astrophysical Virtual Observatory (GAVO). This work benefited from using AWOB, the Astronomer's Workbench. AWOB is a project of the Max Planck Digital Library (MPDL) with the Max Planck Institute for Astrophysics (MPA) and the Max Planck Institute for Extraterrestrial Physics (MPE) as partners.
\end{acknowledgements}


\begin{thebibliography}{90}
\expandafter\ifx\csname natexlab\endcsname\relax\def\natexlab#1{#1}\fi

\bibitem[Adelberger et al.(2005)]{adelberger05} Adelberger, K.~L., Steidel, C.~C., Pettini, M., et al.\ 2005, \apj, 619, 697 


\bibitem[Andreon(2008)]{andreon08} Andreon, S.\ 2008, \mnras, 386, 1045 

\bibitem[Andreon(2010)]{andreon10} Andreon, S.\ 2010, \mnras, 
407, 263 

\bibitem[Angulo et al.(2012)]{angulo12} Angulo, R.~E., Springel, 
V., White, S.~D.~M., et al.\ 2012, \mnras, 425, 2722 

\bibitem[Bah{\'e} et al.(2012)]{bahe12} Bah{\'e}, Y.~M., 
McCarthy, I.~G., \& King, L.~J.\ 2012, \mnras, 421, 1073 

\bibitem[Bardeen et al.(1986)]{bardeen86} Bardeen, J.~M., Bond, 
J.~R., Kaiser, N., \& Szalay, A.~S.\ 1986, \apj, 304, 15 

\bibitem[Bauer et al.(2011)]{bauer11} Bauer, A.~E., Gr{\"u}tzbauch, R., J{\o}rgensen, I., Varela, J., \& Bergmann, M.\ 2011, \mnras, 411, 2009 


\bibitem[Bertone \& Conselice(2009)]{bertone09} Bertone, S., \& Conselice, C.~J.\ 2009, \mnras, 396, 2345 

\bibitem[Blakeslee et al.(2003)]{blakeslee03} Blakeslee, J.~P., Franx, M., Postman, M., et al.\ 2003, \apjl, 596, L143 

\bibitem[Blain et al.(2004)]{blain04} Blain, A.~W., Chapman, S.~C., Smail, I., \& Ivison, R.\ 2004, \apj, 611, 725 

\bibitem[Capak et al.(2011)]{capak11} Capak, P.~L., Riechers, 
D., Scoville, N.~Z., et al.\ 2011, \nat, 470, 233 

\bibitem[Carroll et al.(1992)]{carroll92} Carroll, S.~M., Press, W.~H., \& Turner, E.~L.\ 1992, \araa, 30, 499 


\bibitem[Chiaberge et al.(2010)]{Chiaberge10} Chiaberge, M., 
Capetti, A., Macchetto, F.~D., et al.\ 2010, \apjl, 710, L107

\bibitem[Cohn et al.(2007)]{cohn07} Cohn, J.~D., Evrard, 
A.~E., White, M., Croton, D., \& Ellingson, E.\ 2007, \mnras, 382, 1738 

\bibitem[Colless et al.(2001)]{colless01} Colless, M., Dalton, 
G., Maddox, S., et al.\ 2001, \mnras, 328, 1039 


\bibitem[Croft et al.(2005)]{croft05} Croft, S., Kurk, J., van 
Breugel, W., et al.\ 2005, \aj, 130, 867 

\bibitem[Croton et al.(2006)]{croton06} Croton, D.~J., Springel, 
V., White, S.~D.~M., et al.\ 2006, \mnras, 365, 11 

\bibitem[De Lucia 
\& Blaizot(2007)]{DLB07} De Lucia, G., \& Blaizot, J.\ 2007, \mnras, 375, 2 

\bibitem[Digby-North et al.(2010)]{digby-north10} Digby-North, J.~A., 
Nandra, K., Laird, E.~S., et al.\ 2010, \mnras, 407, 846 

\bibitem[Dressler(1980)]{dressler80} Dressler, A.\ 1980, \apj, 236, 351 

\bibitem[Ebeling et al.(2001)]{ebeling01} Ebeling, H., Edge, 
A.~C., \& Henry, J.~P.\ 2001, \apj, 553, 668 


\bibitem[Elbaz et al.(2007)]{elbaz07} Elbaz, D., Daddi, E., Le Borgne, D., et al.\ 2007, \aap, 468, 33 

\bibitem[Fanidakis et al.(2012)]{fanidakis12} Fanidakis, N., Baugh, 
C.~M., Benson, A.~J., et al.\ 2012, \mnras, 419, 2797 

\bibitem[Fassbender et al.(2011a)]{fassbender11a} Fassbender, R., B{\"o}hringer, H., Nastasi, A., et al.\ 2011, New Journal of Physics, 13, 125014 

\bibitem[Fassbender et al.(2011b)]{fassbender11b} Fassbender, R., Nastasi, A., B{\"o}hringer, H., et al.\ 2011, \aap, 527, L10

\bibitem[Foley et al.(2011)]{foley11} Foley, R.~J., Andersson, K., Bazin, G., et al.\ 2011, \apj, 731, 86 

\bibitem[Galametz et al.(2010)]{galametz10} Galametz, A., Vernet, J., De Breuck, C., et al.\ 2010, \aap, 522, A58 

\bibitem[Gawiser et al.(2007)]{gawiser07} Gawiser, E., Francke, 
H., Lai, K., et al.\ 2007, \apj, 671, 278

\bibitem[Genel et al.(2008)]{genel08} Genel, S., Genzel, R., 
Bouch{\'e}, N., et al.\ 2008, \apj, 688, 789 

\bibitem[Genel et al.(2009)]{genel09} Genel, S., Genzel, R., 
Bouch{\'e}, N., Naab, T., \& Sternberg, A.\ 2009, \apj, 701, 2002 


\bibitem[Giodini et al.(2009)]{giodini09} Giodini, S., Pierini, 
D., Finoguenov, A., et al.\ 2009, \apj, 703, 982 

\bibitem[Gladders \& Yee(2005)]{gladders05} Gladders, M.~D., \& Yee, H.~K.~C.\ 2005, \apjs, 157, 1 

\bibitem[Gonzalez et al.(2005)]{gonzalez05} Gonzalez, A.~H., Tran,
K.-V.~H., Conbere, M.~N., \& Zaritsky, D.\ 2005, \apjl, 624, L73

\bibitem[Gonzalez et al.(2007)]{gonzalez07} Gonzalez, A.~H., 
Zaritsky, D., \& Zabludoff, A.~I.\ 2007, \apj, 666, 147 

\bibitem[Goto et al.(2003)]{goto03} Goto, T., Yamauchi, C., 
Fujita, Y., et al.\ 2003, \mnras, 346, 601 

\bibitem[Goto et al.(2008)]{goto08} Goto, T., Hanami, H., Im, M., et al.\ 2008, \pasj, 60, 531 


\bibitem[Gr{\"u}tzbauch et al.(2011)]{2011MNRAS.418..938G} Gr{\"u}tzbauch, 
R., Conselice, C.~J., Bauer, A.~E., et al.\ 2011, \mnras, 418, 938


\bibitem[Guo \& White(2009)]{guo09} Guo, Q., \& White, S.~D.~M.\ 2009, \mnras, 396, 39 

\bibitem[Guo et al.(2011)]{guo11} Guo, Q., White, S., 
Boylan-Kolchin, M., et al.\ 2011, \mnras, 413, 101 



\bibitem[Guo et al.(2013)]{guo13} Guo, Q., White, S., Angulo, 
R.~E., et al.\ 2013, \mnras, 428, 1351 

\bibitem[Hatch et al.(2011)]{hatch11a} Hatch, N.~A., De Breuck, 
C., Galametz, A., et al.\ 2011, \mnras, 410, 1537 

\bibitem[Hatch et al.(2011)]{hatch11b} Hatch, N.~A., Kurk, 
J.~D., Pentericci, L., et al.\ 2011, \mnras, 415, 2993 

\bibitem[Henriques et al.(2012)]{henriques12} Henriques, B.~M.~B., White, S.~D.~M., Lemson, G., et al.\ 2012, \mnras, 421, 2904 

\bibitem[Henry et al.(2010)]{henry10} Henry, J.~P., Salvato, 
M., Finoguenov, A., et al.\ 2010, \apj, 725, 615 

\bibitem[Horesh et al.(2011)]{horesh11} Horesh, A., Maoz, D., 
Hilbert, S., \& Bartelmann, M.\ 2011, \mnras, 418, 54 

\bibitem[Hayashi et al.(2012)]{hayashi12} Hayashi, M., Kodama, 
T., Tadaki, K.-i., Koyama, Y., \& Tanaka, I.\ 2012, \apj, 757, 15 

\bibitem[Hopkins \& Beacom(2006)]{hopkins06} Hopkins, A.~M., \& Beacom, J.~F.\ 2006, \apj, 651, 142 


\bibitem[Ivison et al.(2013)]{ivison13} Ivison, R.~J., Swinbank, 
A.~M., Smail, I., et al.\ 2013, arXiv:1302.4436

\bibitem[Jose et al.(2013)]{jose13} Jose, C., Srianand, R., 
\& Subramanian, K.\ 2013, arXiv:1304.7458 



\bibitem[Komatsu et al.(2011)]{komatsu11} Komatsu, E., Smith, K.~M., Dunkley, J., et al.\ 2011, \apjs, 192, 18 

\bibitem[Kova{\v c} et al.(2007)]{kovac07} Kova{\v c}, K., Somerville, R.~S., Rhoads, J.~E., Malhotra, S., \& Wang, J.\ 2007, \apj, 668, 15 

\bibitem[Koyama et al.(2013)]{koyama13a} Koyama, Y., Kodama, T., 
Tadaki, K.-i., et al.\ 2013, \mnras, 428, 1551 


\bibitem[Kuiper et al.(2011)]{kuiper11} Kuiper, E., Hatch, 
N.~A., Venemans, B.~P., et al.\ 2011, \mnras, 417, 1088 

\bibitem[Kulas et al.(2013)]{kulas13} Kulas, K.~R., McLean, 
I.~S., Shapley, A.~E., et al.\ 2013, \apj, 774, 130 

\bibitem[Kurk et al.(2000)]{kurk00} Kurk, J.~D., R{\"o}ttgering, H.~J.~A., Pentericci, L., et al.\ 2000, \aap, 358, L1 

\bibitem[Kurk et al.(2004)]{kurk04a} Kurk, J.~D., Pentericci, L., R{\"o}ttgering, H.~J.~A., \& Miley, G.~K.\ 2004, \aap, 428, 793 

\bibitem[Kurk et al.(2004)]{kurk04b} Kurk, J.~D., Pentericci, L., Overzier, R.~A., R{\"o}ttgering, H.~J.~A., \& Miley, G.~K.\ 2004, \aap, 428, 817


\bibitem[Lehmer et al.(2009)]{lehmer09} Lehmer, B.~D., 
Alexander, D.~M., Geach, J.~E., et al.\ 2009, \apj, 691, 687 

\bibitem[Lemson \& Virgo Consortium(2006)]{lemson06} Lemson, G., \& Virgo Consortium, t.\ 2006, arXiv:astro-ph/0608019 

\bibitem[Lidman et al.(2008)]{lidman08} Lidman, C., Rosati, P., Tanaka, M., et al.\ 2008, \aap, 489, 981 

\bibitem[Martini et al.(2013)]{martini13} Martini, P., Miller, 
E.~D., Brodwin, M., et al.\ 2013, \apj, 768, 1 

\bibitem[Matsuda et al.(2005)]{matsuda05} Matsuda, Y., Yamada, 
T., Hayashino, T., et al.\ 2005, \apjl, 634, L125 

\bibitem[Matsuda et al.(2009)]{2009MNRAS.400L..66M} Matsuda, Y., Nakamura, 
Y., Morimoto, N., et al.\ 2009, \mnras, 400, L66 

\bibitem[Matsuda et al.(2012)]{matsuda12} Matsuda, Y., Yamada, 
T., Hayashino, T., et al.\ 2012, \mnras, 425, 878 

\bibitem[Mei et al.(2006)]{mei06} Mei, S., Holden, B.~P., Blakeslee, J.~P., et al.\ 2006, \apj, 644, 759

\bibitem[Menanteau et al.(2012)]{menanteau12} Menanteau, F., Sif{\'o}n, C., Barrientos, L.~F., et al.\ 2012, arXiv:1210.4048 

\bibitem[Merson et al.(2013)]{merson13} Merson, A.~I., Baugh, C.~M., Helly, J.~C., et al.\ 2013, \mnras, 429, 556 


\bibitem[Miley et al.(2004)]{miley04} Miley, G.~K., Overzier, 
R.~A., Tsvetanov, Z.~I., et al.\ 2004, \nat, 427, 47 

\bibitem[Mo \& White(1996)]{mo96} Mo, H.~J., \& White, S.~D.~M.\ 1996, \mnras, 282, 347

\bibitem[Mostardi et al.(2013)]{mostardi13} Mostardi, R.~E., Shapley, A.~E., Nestor, D.~B., Steidel, C.~C., \& Reddy, N.~A.\ 2013, arXiv:1306.1535 



\bibitem[Olsen et al.(2007)]{olsen07} Olsen, L.~F., Benoist, C., Cappi, A., et al.\ 2007, \aap, 461, 81 



\bibitem[Overzier et al.(2006)]{overzier06b} Overzier, R.~A., Miley, G.~K., Bouwens, R.~J., et al.\ 2006, \apj, 637, 58 

\bibitem[Overzier et al.(2008)]{overzier08} Overzier, R.~A., 
Bouwens, R.~J., Cross, N.~J.~G., et al.\ 2008, \apj, 673, 143 

\bibitem[Overzier et al.(2009)]{overzier09} Overzier, R.~A., Guo, 
Q., Kauffmann, G., et al.\ 2009, \mnras, 394, 577 


\bibitem[Overzier et al.(2013)]{overzier13} Overzier, R., Lemson, 
G., Angulo, R.~E., et al.\ 2013, \mnras, 428, 778 

\bibitem[Ouchi et al.(2005)]{ouchi05} Ouchi, M., Shimasaku, K., 
Akiyama, M., et al.\ 2005, \apjl, 620, L1 

\bibitem[Ouchi et al.(2005)]{ouchi05b} Ouchi, M., Hamana, T., 
Shimasaku, K., et al.\ 2005, \apjl, 635, L117 

\bibitem[Ouchi et al.(2010)]{ouchi10} Ouchi, M., Shimasaku, K., 
Furusawa, H., et al.\ 2010, \apj, 723, 869 

\bibitem[Palunas et al.(2004)]{palunas04} Palunas, P., Teplitz, 
H.~I., Francis, P.~J., Williger, G.~M., 
\& Woodgate, B.~E.\ 2004, \apj, 602, 545 

\bibitem[Papovich et al.(2010)]{papovich10} Papovich, C., Momcheva, I., Willmer, C.~N.~A., et al.\ 2010, \apj, 716, 1503 

\bibitem[Papovich et al.(2012)]{papovich12} Papovich, C., Bassett, 
R., Lotz, J.~M., et al.\ 2012, \apj, 750, 93

\bibitem[Patel et al.(2009)]{patel09} Patel, S.~G., Holden, B.~P., Kelson, D.~D., Illingworth, G.~D., \& Franx, M.\ 2009, \apjl, 705, L67 

\bibitem[Peacock(1999)]{peacock99} Peacock, J.~A.\ 1999, Cosmological Physics, by John A.~Peacock, pp.~704.~ISBN 052141072X.~Cambridge, UK: Cambridge University Press, January 1999., 

\bibitem[Pentericci et al.(2000)]{pentericci00} Pentericci, L., Kurk, J.~D., R{\"o}ttgering, H.~J.~A., et al.\ 2000, \aap, 361, L25 

\bibitem[Pentericci et al.(2002)]{pentericci02} Pentericci, L., Kurk, J.~D., Carilli, C.~L., et al.\ 2002, \aap, 396, 109 

\bibitem[Poggianti et al.(2008)]{poggianti08} Poggianti, B.~M., 
Desai, V., Finn, R., et al.\ 2008, \apj, 684, 888 

\bibitem[Prescott et al.(2008)]{prescott08} Prescott, M.~K.~M., 
Kashikawa, N., Dey, A., \& Matsuda, Y.\ 2008, \apjl, 678, L77 

\bibitem[Quilis \& Trujillo(2012)]{quilis12} Quilis, V., \& Trujillo, I.\ 2012, \apjl, 752, L19 

\bibitem[Rettura et al.(2010)]{rettura10} Rettura, A., Rosati, P., Nonino, M., et al.\ 2010, \apj, 709, 512

\bibitem[Rettura et al.(2011)]{rettura11} Rettura, A., Mei, S., Stanford, S.~A., et al.\ 2011, \apj, 732, 94 

\bibitem[Santos et al.(2011)]{santos11} Santos, J.~S., Fassbender, R., Nastasi, A., et al.\ 2011, \aap, 531, L15 

\bibitem[Saro et al.(2009)]{saro09} Saro, A., Borgani, S.,
Tornatore, L., et al.\ 2009, \mnras, 392, 795


\bibitem[Shimasaku et al.(2003)]{shimasaku03} Shimasaku, K., Ouchi, 
M., Okamura, S., et al.\ 2003, \apjl, 586, L111 

\bibitem[Shimasaku et al.(2004)]{shimasaku04} Shimasaku, K., 
Hayashino, T., Matsuda, Y., et al.\ 2004, \apjl, 605, L93 


\bibitem[Springel et al.(2001)]{springel01} Springel, V., White, 
S.~D.~M., Tormen, G., \& Kauffmann, G.\ 2001, \mnras, 328, 726 

\bibitem[Spergel et al.(2003)]{spergel03} Spergel, D.~N., Verde, 
L., Peiris, H.~V., et al.\ 2003, \apjs, 148, 175 

\bibitem[Spitler et al.(2012)]{spitler12} Spitler, L.~R., 
Labb{\'e}, I., Glazebrook, K., et al.\ 2012, \apjl, 748, L21 

\bibitem[Springel et al.(2005)]{springel05} Springel, V., White, S.~D.~M., Jenkins, A., et al.\ 2005, \nat, 435, 629 

\bibitem[Steidel et al.(1998)]{steidel98} Steidel, C.~C., Adelberger, K.~L., Dickinson, M., et al.\ 1998, \apj, 492, 428

\bibitem[Steidel et al.(2000)]{steidel00} Steidel, C.~C., Adelberger, K.~L., Shapley, A.~E., et al.\ 2000, \apj, 532, 170 

\bibitem[Steidel et al.(2005)]{steidel05} Steidel, C.~C., Adelberger, K.~L., Shapley, A.~E., et al.\ 2005, \apj, 626, 44

\bibitem[Stevens et al.(2003)]{stevens03} Stevens, J.~A., Ivison, R.~J., Dunlop, J.~S., et al.\ 2003, \nat, 425, 264 

\bibitem[Suwa et al.(2006)]{suwa06} Suwa, T., Habe, A., 
\& Yoshikawa, K.\ 2006, \apjl, 646, L5 


\bibitem[Tanaka et al.(2010)]{tanaka10} Tanaka, M., Finoguenov, 
A., \& Ueda, Y.\ 2010, \apjl, 716, L152 

\bibitem[Tanaka et al.(2011)]{tanaka11} Tanaka, I., Breuck, 
C.~D., Kurk, J.~D., et al.\ 2011, \pasj, 63, 415 

\bibitem[Toshikawa et al.(2012)]{toshikawa12} Toshikawa, J., Kashikawa, N., Ota, K., et al.\ 2012, \apj, 750, 137 

\bibitem[Tran et al.(2010)]{tran10} Tran, K.-V.~H., Papovich, 
C., Saintonge, A., et al.\ 2010, \apjl, 719, L126 

\bibitem[Trenti et al.(2012)]{trenti12} Trenti, M., Bradley, 
L.~D., Stiavelli, M., et al.\ 2012, \apj, 746, 55 

\bibitem[Utsumi et al.(2010)]{utsumi10} Utsumi, Y., Goto, T., 
Kashikawa, N., et al.\ 2010, \apj, 721, 1680 

\bibitem[Venemans et al.(2002)]{venemans02} Venemans, B.~P., Kurk, 
J.~D., Miley, G.~K., et al.\ 2002, \apjl, 569, L11 

\bibitem[Venemans et al.(2004)]{venemans04} Venemans, B.~P., R{\"o}ttgering, H.~J.~A., Overzier, R.~A., et al.\ 2004, \aap, 424, L17 


\bibitem[Venemans et al.(2005)]{venemans05b} Venemans, B.~P., R{\"o}ttgering, H.~J.~A., Miley, G.~K., et al.\ 2005, \aap, 431, 793 

\bibitem[Venemans et al.(2007)]{venemans07} Venemans, B.~P., R{\"o}ttgering, H.~J.~A., Miley, G.~K., et al.\ 2007, \aap, 461, 823 

\bibitem[Wilson et al.(2009)]{wilson09} Wilson, G., Muzzin, A., Yee, H.~K.~C., et al.\ 2009, \apj, 698, 1943 

\bibitem[Wu et al.(2013)]{wu13} Wu, H.-Y., Hahn, O., Wechsler, R.~H., Mao, Y.-Y., \& Behroozi, P.~S.\ 2013, \apj, 763, 70 

\bibitem[Yamada et al.(2012)]{yamada12} Yamada, T., Nakamura, 
Y., Matsuda, Y., et al.\ 2012, \aj, 143, 79

\bibitem[Yang et al.(2009)]{yang09} Yang, Y., Zabludoff, A., 
Tremonti, C., Eisenstein, D., \& Dav{\'e}, R.\ 2009, \apj, 693, 1579 


\bibitem[Zirm et al.(2005)]{zirm05} Zirm, A.~W., Overzier, 
R.~A., Miley, G.~K., et al.\ 2005, \apj, 630, 68









\end{thebibliography}
\end{document}